\documentclass{aa}
\usepackage{graphicx} 
\usepackage{txfonts}
\usepackage{multirow}
\usepackage{natbib}
\bibpunct{(}{)}{;}{a}{}{,}
\usepackage[colorlinks=true,citecolor=blue, urlcolor=blue, linkcolor=blue,breaklinks=true]{hyperref}
\usepackage{xspace}
\usepackage{amsmath,amssymb}
\usepackage{booktabs}
\usepackage[shortlabels]{enumitem}
\usepackage{upgreek}
\usepackage{siunitx}  
\usepackage{microtype}
\let\linenumbers\relax
\let\nolinenumbers\relax
\usepackage{tabularx}
\usepackage{float}
\usepackage{etoolbox}
\patchcmd{\section}{\clearpage}{}{}{} 

\usepackage{afterpage}   
\newcommand{\logg}{\ensuremath{\log g}\xspace}  

\newcommand{\co}{\ensuremath{\mathrm{C/O}}\xspace}

\newcommand{\Kzz}{\ensuremath{K_{\textrm{zz}}}\xspace}
\newcommand{\vsed}{\ensuremath{v_{\textrm{sed}}}\xspace}
\newcommand{\fsed}{\ensuremath{f_{\textrm{sed}}}\xspace}
\newcommand{\Teff}{\ensuremath{T_{\textrm{eff}}}\xspace}
\newcommand{\taunuage}{\ensuremath{\tau_{\textrm{cloud}}}\xspace}
\newcommand{\fsedun}{\ensuremath{f_{\textrm{sed},\,1}}\xspace}
\newcommand{\fseddeux}{\ensuremath{f_{\textrm{sed},\,2}}\xspace}
\newcommand{\chisqreduced}{\ensuremath{\chi^2_{\rm red}}\xspace}

\newcommand{\Nspec}{\ensuremath{N_{\textrm{spec}}}\xspace}
\newcommand{\kapext}{\ensuremath{\kappa_{\textrm{ext}}}\xspace}
\newcommand{\kapCIA}{\ensuremath{\kappa_{\textrm{CIA}}}\xspace}
\newcommand{\kapRayl}{\ensuremath{\kappa_{\textrm{Rayleigh}}}\xspace}
\newcommand{\kapline}{\ensuremath{\kappa_{\textrm{line}}}\xspace}
\newcommand{\kapaero}{\ensuremath{\kappa_{\textrm{aerosols}}}\xspace}
\newcommand{\Fold}{\ensuremath{F_{\textrm{old}}}\xspace}
\newcommand{\Fnew}{\ensuremath{F_{\textrm{new}}}\xspace}
\newcommand{\FUL}{\ensuremath{F_{\textrm{UL}}}\xspace}
\newcommand{\mubruit}{\ensuremath{\mu_{\textrm{noise}}}\xspace}
\newcommand{\sigUL}{\ensuremath{\sigma_{\textrm{UL}}}\xspace}
\newcommand{\Fcomb}{\ensuremath{F_{\textrm{comb}}}\xspace}
\newcommand{\Ffsedun}{\ensuremath{F_{f\textrm{sed},\,1}}\xspace}
\newcommand{\Ffseddeux}{\ensuremath{F_{f\textrm{sed},\,2}}\xspace}
\newcommand{\taucoal}{\ensuremath{\tau_{\textrm{coal}}}\xspace}
\newcommand{\taucond}{\ensuremath{\tau_{\textrm{cond}}}\xspace}
\newcommand{\taused}{\ensuremath{\tau_{\textrm{sed}}}\xspace}
\newcommand{\taumix}{\ensuremath{\tau_{\textrm{mix}}}\xspace}
\newcommand{\tauadv}{\ensuremath{\tau_{\textrm{adv}}}\xspace}
\newcommand{\taurad}{\ensuremath{\tau_{\textrm{rad}}}\xspace}
\newcommand{\Aobs}{\ensuremath{A_{\textrm{obs}}}\xspace}
\newcommand{\LSun}{\ensuremath{L_{\odot}}\xspace}
\newcommand{\MSun}{\ensuremath{M_{\odot}}\xspace}
\newcommand{\MJup}{\ensuremath{M_{\mathrm{Jup}}}\xspace}
\newcommand{\RJup}{\ensuremath{R_{\mathrm{Jup}}}\xspace}

\newcommand{\mic}{\ensuremath{\upmu\mathrm{m}}\xspace}
\newcommand{\exo}{\texttt{Exo-REM}\xspace}
\newcommand{\exoII}{{\small\textls[-50]{\texttt{Exo-REM~k26}}}\xspace}
\newcommand{\exok}{\texttt{Exo\_k}\xspace}
\newcommand{\formosa}{\texttt{ForMoSA}\xspace}

\newcommand{\jwst}{JWST\xspace}

\newcommand{\vhs}{VHS~1256~b\xspace}

\def\largimgunecol{0.47\textwidth}
\def\largimgdeuxcols{0.97\textwidth}
\usepackage[weather]{ifsym}    
\begin{document}
\title{Next-generation Exo-REM atmospheric models}
\subtitle{Application to \vhs to emulate patchy clouds}

\author{
    Alice~Radcliffe \inst{\ref{lesia}}
    \and Benjamin~Charnay \inst{\ref{lesia},\ref{bordeaux}}
    \and Anne-Marie~Lagrange \inst{\ref{lesia}}
    \and Flavien~Kiefer \inst{\ref{lesia}}
    \and Bruno~B{\'e}zard \inst{\ref{lesia}}
    \and Simon~Petrus \inst{\ref{NASA}}
    \and Paulina~Palma-Bifani \inst{\ref{lesia}}
    \and Matthieu Ravet \inst{\ref{lagrange},\ref{IPAG},\ref{MPIA}}
    \and J{\'e}r{\'e}my~Leconte \inst{\ref{bordeaux}}
    \and Gabriel-Dominique~Marleau \inst{\ref{Duisburg},\ref{Bern},\ref{MPIA}}
    }

\institute{
    LIRA, Observatoire de Paris, Univ. PSL, CNRS, Sorbonne Univ., Univ. Paris Cit\'e, 5 place Jules Janssen, 92195 Meudon, France \label{lesia}
    \and
    Laboratoire d’astrophysique de Bordeaux, Univ. Bordeaux, CNRS, B18N, all\'ee Geoffroy Saint-Hilaire, 33615 Pessac, France \label{bordeaux}
    \and
    NASA-Goddard Space Flight Center, Greenbelt, MD 20771, USA \label{NASA}
    \and
    Laboratoire J.-L. Lagrange, Université Côte d'Azur, Observatoire de la Côte d'Azur, CNRS, 06304 Nice, France\label{lagrange}
    \and
    IPAG, Université Grenoble-Alpes, CNRS, F-38000 Grenoble, France\label{IPAG}
    \and
    Max-Planck-Institut f\"ur Astronomie,
    K\"onigstuhl 17,
    69117 Heidelberg, Germany \label{MPIA}
    \and
    Fakult\"at f\"ur Physik,
    Universit\"at Duisburg--Essen,
    Lotharstra\ss{}e~1,
    47057 Duisburg, Germany \label{Duisburg}
    \and
    Abteilung f\"ur Weltraumforschung und Planetologie,
    Physikalisches Institut, Universit\"at Bern,
    Sidlerstr.~5,
    3012 Bern, Switzerland \label{Bern}
}
\date{Received 19 January 2026 / Accepted 21 May 2026}

\abstract{Condensate clouds are a defining feature of brown dwarf and exoplanet atmospheres, producing a broad range of colours on the colour–magnitude diagram (CMD) and giving rise to spectral features such as the distinct $\sim$10\,$\mic$ spectral imprint, a prominent diagnostic for silicate clouds. Cloud cover is likely to be heterogeneous in many objects, with observed rotational variability providing key evidence for the presence of thick and thin cloud regions rotating in and out of view. Yet current one-dimensional (1D) atmosphere models, often lacking any parameter to tune cloud optical thickness, typically fail to reproduce the spectra of highly cloudy substellar objects, especially those with complex cloud structures. Our goal is to address these limitations by upgrading the \exo atmosphere model, and by devising a more nuanced approach to better describe heterogeneous cloud cover with pre-computed 1D grids. Here, we present new self-consistent low- ($R = 500$) and medium-resolution ($R = \textrm{10,000}$) \exo grids, hereafter \exoII, featuring critical updates: (1)~the incorporation of a cloud sedimentation parameter, \fsed, to govern cloud opacity, thereby enabling even the reddest of objects to be accessed on a CMD, revealing a trend of decreasing \fsed along the L--T transition (2)~the substantial update of molecular opacities and abundances used, including new experimentally validated alkali line lists, and (3)~the implementation of strict convergence criteria that entirely avoid unstable model solutions. 
Correcting an erroneous $\text{CH}_3\text{D}$ abundance leads to marked spectral changes for low-\Teff\ (methane-rich) objects. As a consequence, applying \exoII to the cool GJ~504\,b leads to a revision of its parameters ($\Teff = 473^{+14}_{-12}$\,K, $\logg = 4.0\pm 0.1$\,dex). For the notoriously variable \vhs, a two-column \exoII framework that emulates cloud heterogeneities achieves a significantly improved global fit over a single 1D model. Here, a $\sim$ 60-40\% split of thick and thin clouds best describes its atmosphere, further confirming the presence of patchy clouds. In particular, this reproduces the strong $10\,\mic$ silicate absorption in the MIRI/MRS (\jwst) data of \vhs, where 1D grids had previously failed, owing to the formerly unexplored low-\fsed regime in the new model. The addition of the \fsed dimension, as well as the two-column approach for heterogeneous cloud distributions, prove vital for accurately characterizing cloudy sub-stellar objects.}


\maketitle

\section{Introduction}
\label{section:introduction}

Condensate clouds are key elements in the atmospheres of brown dwarfs (BDs) \citep{Kirkpatrick2005, Marley2015} and exoplanets \citep{Madhusudhan2019}, where they have a major impact on their emitted or transmitted spectra, and are expected to have a pronounced effect on future  reflected light observations with next-generation instruments such as the Extremely Large Telescope (ELT) and the Habitable Worlds Observatory (HWO). Changes in cloud altitude with effective temperature are generally understood to be the main drivers of spectral type and colour evolution in BDs (e.g.\ \citealt{Cushing2008,Saumon2008,Charnay2018}): the L--T transition marks a sudden blue shift in colour over a small temperature range of a few hundred kelvin that occurs, as the effective temperature decreases, when silicate and iron condensate clouds begin to sink lower in the atmosphere, rendering it optically thinner at short wavelengths.

To address the challenge of modelling cloud formation and evolution, various 1D atmosphere models adopt different methodological approaches, varying in their complexity and incorporation of physical processes, as well as the range of parameters they explore. \texttt{DRIFT-PHOENIX} \citep{Helling2008, Witte2009, Witte2011} includes a wide variety of cloud species (Fe, SiO$_2$, TiO$_2$, Al$_2$O$_3$, MgO, MgSiO$_3$, and Mg$_2$SiO$_4$) in its non-equilibrium cloud model, but does not take coagulation into account. \texttt{ATMO} \citep{Tremblin2015, phillips2020} is a cloudless model that focuses on the CO/CH$_4$ and N$_2$/NH$_3$ non-equilibrium reactions that induce fingering convection, providing an alternative hypothesis for the reddening phenomenon which excludes the presence of condensate clouds. However, this approach is debated in \citet{Leconte2018}, and the assumption of cloud-free conditions is now contradicted by many unambiguous detections of silicate cloud features \citep{Surez2022, Miles2023, Hoch2025, Mollire2025}. \texttt{BT-Settl} \citep{Allard2012} is a model that estimates the abundance and size distributions of cloud particles for 55 types of solids, with non-equilibrium chemistry included for several species. \cite{Lacy2023} incorporate disequilibrium chemistry and water clouds in their self-consistent models of Y-dwarf atmospheres. Finally, the \texttt{Sonora-Diamondback} model \citep{Morley2024} belongs to the \texttt{Sonora} family, along with the \texttt{Sonora-Bobcat} \citep{Marley2021}, \texttt{Sonora-Cholla} \citep{Karalidi2021} and \texttt{Sonora Elf-Owl} \citep{Mukherjee2024,Wogan2025}, and is the only \texttt{Sonora} model that includes clouds (following the approach in \citealt{Ackerman2001}), parametrizing them by exploring their vertical extent and opacity. The ranges of parameters explored in each model are summarised in Table~\ref{tab: models}.

\begin{table}[t]
\centering
\caption{Comparison of substellar atmosphere model parameter ranges}
\label{tab: models}
\footnotesize
\setlength{\tabcolsep}{3pt}
\begin{tabular}{l ccccc}
\hline\hline
Model & \Teff & \logg & [M/H] & C/O\tablefootmark{a} & \fsed\tablefootmark{b} \\
      & [K]   & [dex] & [dex] &   &     \\
\hline
\texttt{ATMO}$^1$ (cloud-free)     & 200--3000  & 2.5--5.5 & $-0.6$--$0.6$ & 0.3--0.7 & $\cdots$ \\
\texttt{DRIFT-PHOENIX}$^2$ & 1000--1300 & 3.0--6.0 & $-0.6$--$0.3$ & 0.55 & \FilledSunCloud \\ 
\texttt{BT-Settl CIFIST}$^3$  & 1200--7000 & 2.5--5.5 & 0.0 & 0.55 & \FilledSunCloud \\ 
\texttt{Lacy \&\ Burrows}$^4$
& 200--450 & 3.75--5.0 & $-0.5$--$0.5$  & $\cdots$ & \FilledSunCloud \\  
\texttt{Sonora Db}$^5$ & 900--2400  & 3.5--5.5 & $-0.5$--$0.5$ & 0.458 & 1--8 \\  
\textls[-50]{\texttt{Exo-REM k26}}$^6$ & 200--2000 & 3.0--5.0 & $-0.5$--$2.0$ & 0.1--0.8 & 0.5--9 \\
\hline
\end{tabular}
\tablefoot{
{References}:
(1)~\citet{Tremblin2015, phillips2020}. 
(2)~\citet{Helling2008}.
(3)~\citet{Allard2012}.
(4)~\citet{Lacy2023}.
(5)~\texttt{Db}\,=\,\texttt{Diamondback}; \citet{Morley2024}.
(6)~This work.\\
\tablefoottext{a}{Solar: $\textrm{C/O} = 0.458$ in \citet{Lodders2010}, 0.55 in \citet{Asplund2009} and \citet{Caffau2011}, 0.59 in \citet{Asplund2021}.
}\\
\tablefoottext{b}{The symbol \FilledSunCloud{} marks models with clouds not parametrised by \fsed. Models often also have a cloud-free version.
}
}
\end{table}

Our Exoplanet Radiative-convective Equilibrium Model (\exo) is an atmosphere model that was initially developed to interpret upcoming photometric and spectral measurements from the new generation of instruments such as VLT/SPHERE in order to characterize the atmospheres of young Jupiters \citep{Baudino2015}. Subsequently a more elaborate self-consistent cloud model was developed \citep{Charnay2018}, which accounted for both absorption and scattering of thermal radiation. \exo has since been adapted in \citet{Blain2021} for irradiated planets so as to include transmission spectroscopy. While the existing \exo grids manage to, with a simple microphysics cloud model including iron and forsterite condensates, follow the general trend of the L--T transition on a CMD, the lack of a free parameter for cloud vertical distribution prevents them from reproducing the whole diversity of colours of field BDs and young giant planets (YGPs). Here, we will detail the addition of a parameter designed to adjust the cloud vertical extent and opacity, namely, the sedimentation rate (\fsed) parameter. This has previously been incorporated in modelling BDs vertical profiles and evolution in CMDs \citep{Ackerman2001, Saumon2008} but has only recently been added as an additional dimension in one other atmosphere model (\texttt{Sonora-Diamondback}; \citealt{Morley2024}).

Although the fine-tuning of condensate cloud optical thickness in atmosphere models is essential, it nonetheless assumes uniform cloud coverage over the whole surface of the object and thus does not capture the complexities of objects displaying heterogeneous distributions of clouds. Indeed, time-variations in flux have been widely observed in BDs, such as \vhs \citep{Zhou2020} and many others \citep{Artigau2009, Radigan2012, Radigan2014, Metchev2015, Buenzli2015, Vos2017, Eriksson2019, Zhou2018, Bowler2020, Vos2020, Tannock2021, Vos2022, Biller2024, Nasedkin2025}, and in planetary mass companions such as 2MASS~1207\,b and 2M1207\,b (\citealp{Zhou2016, Adams2025}), strongly suggesting the existence of large-scale atmospheric heterogeneities, most likely in the form of patchy clouds \citep{Marley2010, Apai2013, Morley2014a}. As it rotates, a patchy object would show thicker and thinner cloud zones, rapidly exposing different atmospheric depths and temperatures, leading to periodic light curve variations. To corroborate these observations, global climate models (GCMs) have predicted heterogeneous cloud cover as a natural characteristic in BDs: cloud radiative effects trigger atmospheric convection, giving rise to spatial and temporal changes consistent with the variability seen across the L--T transition \citep{Artigau2018, McCarthy2025, Teinturier2026}. While a thick zonal cloud band is predicted to populate the equator for Coriolis-dominated fast rotators, longitudonal variations are the culprits for the widely-observed time variability. These could come in the shape of zonal waves, storms \citep{Tan2025} or spots \citep{Zhou2022} analogous to the Jupiter's Great Red Spot. However, the physical origins of such structures on directly imaged objects, that have different dominating processes in their atmosphere dynamics \citep{Tan2017}, remain elusive.

Fitting observed spectra of complex patchy objects is a current challenge in atmospheric modelling: while standard 1D atmospheric models, with their assumption of a homogeneous atmosphere, are insufficient, sophisticated three-dimensional GCMs are computationally costly and it is, as it stands, unrealistic to create large grids of these aimed at fitting observations. Intermediate approaches consisting in combining two 1D atmospheric columns differing in cloud properties have successfully been applied in retrieval (e.g.\ \citealt{Vos2023, Zhang2025}, and \citealt{Mollire2025}), and forward modelling (e.g.\ \citealt{Marley2010} and \citealt{Morley2014b, Morley2014a}) frameworks, where in both cases a single thermal profile for the whole atmosphere was converged to. Here we propose an approach consisting in combining two 1D \exoII \footnote{The name we have chosen for this generation of \exo grids includes a `k' -- chosen as a nod to \texttt{Exo\_k} -- and `26`, referring to the year.} atmospheric columns, with differing cloud optical thicknesses and thermal profiles, in a forward-modelling framework, to describe objects with patchy cloud cover without adding parameters to the pre-computed grid. 

\vhs constitutes a prime test case for this method, since time-monitoring characterises it as having one of the highest rotationally modulated variabilities of substellar objects to date: from two epochs (a 9\,h HST 2018 time series \citep{Bowler2020} and a 42\,h HST 2020 time series \citep{Zhou2022}), its peak-to-peak amplitude is estimated at $37.6\pm2.2\%$ at 1.27\,\mic. Additionally, a 38\,h \textit{Spitzer}/IRAC (Infrared Array Camera) observation measured a variability of $5.76\pm 0.04\%$ at 4.5\,\mic \citep{Zhou2020}. The presence of thick silicate clouds was confirmed when it was observed by the JWST High Contrast Early-release Science Programme \citep{Hinkley2022} with a combination of NIRSpec and MIRI \citep{Miles2023}, and showed a strong characteristic silicate absorption feature in the spectrum at $\sim$10\,\mic. Attempting to converge to a single 1D model is insufficient due to its thick patchy clouds: \citet{Petrus2024} and \citet{Lueber2024} demonstrated that none of the existing grids of pre-computed 1D models (\texttt{Sonora}, \texttt{ATMO}, \texttt{Exo-REM}, \texttt{BT-Settl}, \texttt{DRIFT-PHOENIX}) were able to produce an adequate fit for the JWST spectrum, most dramatically at the silicate feature. \citet{Tan2025}, with their GCM, demonstrated that the complex light curve modulations of \vhs could be explained by cloud heterogeneities in the shape of a persisting dust storm; however, with their lack of disequilibrium chemistry and refined grid, they were not able to produce a satisfactory fit of the molecular absorptions in the JWST spectrum. 

We present in Section~\ref{section: model description} the most consequential updates to the \exo model, detailing the addition of an \fsed parameter, updated line lists and abundances, and the treatment of convergence issues in the code. We bring to light in Section~\ref{section: results} the most notable changes in the \exo spectra, and use them on GJ~504\,b to derive more accurate parameter values, and on \vhs to stress-test the added \fsed dimension while using a two-column forward-modelling approach for patchiness. In Section~\ref{section: discussion} we discuss the implications of the updates in \exo, and the limits to which we may deduce the atmospheric structure of highly cloudy objects. In Section~\ref{section: conclusions} we summarise the work, as well as discuss the upcoming upgrade of \exo to high spectral resolution.

\section{Model description}
\label{section: model description}

\subsection{The radiative-convective equillibrium model: Exo-REM}
\label{subsection: exorem}
\exo solves for radiative-convective equilibrium, assuming that the total flux is conserved at each vertical level, the total flux being comprised of radiative and convective flux. For directly imaged objects we neglect the radiation from the exoplanet's host star. The model is parametrized by the following input parameters: the surface gravity \logg of the object (at 1 bar), its effective temperature \Teff, metallicity [M/H], and the carbon-to-oxygen ratio C/O. The external structure is simply assumed to abide by the ideal gas equations, linking the pressure, density, and temperature.

An important source of opacity is the collision-induced absorption (\kapCIA) from H$_2$ molecules (H$_2$--H$_2$) along with that of H$_2$--He and, in some cases (high metallicity and thus H$_2$O abundance), H$_2$O--H$_2$O. We modelled it using data for the three above cases from \texttt{HITRAN} \citep{Karman2019}. We also include Rayleigh scattering \kapRayl, although it is not a major contributor in the wavelength range modelled here ($\lambda> 1$\,\mic). The line absorption, \kapline, comes from the ro-vibrational bands originating from eleven different molecules, and the presence of resonant lines from Na and K. Finally, the last source of opacity comes from clouds of condensates (\kapaero), so that the total extinction \kapext can be described as the following sum:
\begin{equation}
    \label{eq: extinction coefficient}
    \kapext = \kapCIA  + \kapRayl + \kapline + \kapaero.
\end{equation}
For the relatively cool low-mass objects (200\,K $<$ \Teff $<$ 2000\,K) considered in this model, we include $\Nspec=13$ species (molecules and atoms) as opacity sources in \kapline:
H$_2$O, CH$_4$, CO, CO$_2$, H$_2$S, HCN, K, Na, NH$_3$, PH$_3$, TiO, VO, and FeH (Table~\ref{tab:isotopologues} lists all isotopologues included). Their vertical volume mixing ratio (VMR) profiles are determined by calculating the chemical abundances level by level, starting from the deepest level, and moving upwards to the top of the atmosphere. Examples are shown in Fig.~\ref{fig:vmrs}. These profiles have 49 pressure levels with corresponding temperatures, the top of the atmosphere being the level with the lowest pressure value. Other than the \Nspec species listed above, the simulated atmosphere contains only He and H$_2$, so that, for each pressure level, $\sum_{i=1}^{\Nspec+2}\mathrm{VMR}_i=1$ with an He/H (including hydrogen-bearing species) number ratio $\alpha=0.0839$, from the solar abundance according to \citet{Asplund2021}.

\subsection{Line opacities}

To incorporate \kapline from Eq.~\ref{eq: extinction coefficient} into the model, we need line opacities for each of the thirteen included species, at a range of temperatures and pressures; these can be calculated from available line lists, as provided by, for example, \texttt{HiTEMP} \citep{Rothman2010}, \texttt{TheoReTS} \citep{Rey2016} or \texttt{Exomol} \citep{Tennyson2024}. A Voigt profile is assumed for the molecular lineshape up to a certain distance of line centre, beyond which a sub-Lorentzian profile is applied. The treatment of the wings for atomic resonant lines is particularly important as it can lead to important differences in the final spectrum, as we will see in the case of the alkali atoms included in the model. In addition, the isotopologue abundances must be specified in the instances where opacity data is available for several isotopologues, in which case we include them with natural abundances as listed in \citet{Asplund2021}. In practice, this is done by a linear combination of the $R = 10^6$ cross section data for each isotopologue of a molecule, weighted by their natural abundances listed in Table~\ref{tab:isotopologues} that we computed from Table~\ref{tab:isotopes}, combining them into one single cross section file, as illustrated in Fig.~\ref{fig: ch4_iso} for methane. This is done using \exok\footnote{\url{https://perso.astrophy.u-bordeaux.fr/~jleconte/exo_k-doc/index.html}}, a python code developed to handle cross sections and opacities \citep{Leconte2021}.

\subsubsection{Updated line lists and isotopologue abundances}
\label{subsubsection: line lists}
Recognising a need to update certain line lists and isotopologue abundances used in \texttt{Exo-REM} -- there being advances in their computation as well as the number of isotopologues available -- we revisited them by adding all of the isotopologues for which there was available opacity data. Previously in \exo, solely the main isotopologue was included for each molecule, except in the case of CH$_4$, where $^{12}$CH$_4$, $^{13}$CH$_4$, and CH$_3$D were included, as well as the case of NH$_3$ where $^{14}$NH$_3$ and $^{15}$NH$_3$ were included with a $^{14}$N/$^{15}$N ratio of 500 \citep{Blain2021, Fri2015}. 

Here we have used up-to-date opacities that have been calculated, from many different sources, by \citet{Mollire2019} and can be found in the \texttt{petitRADTRANS} documentation\footnote{\url{https://petitradtrans.readthedocs.io}}.
All of the details are listed in Table~\ref{tab:isotopologues}, and below we discuss the most notable changes.

\begin{figure}[t]
    \centering
    \includegraphics[width=\largimgunecol]{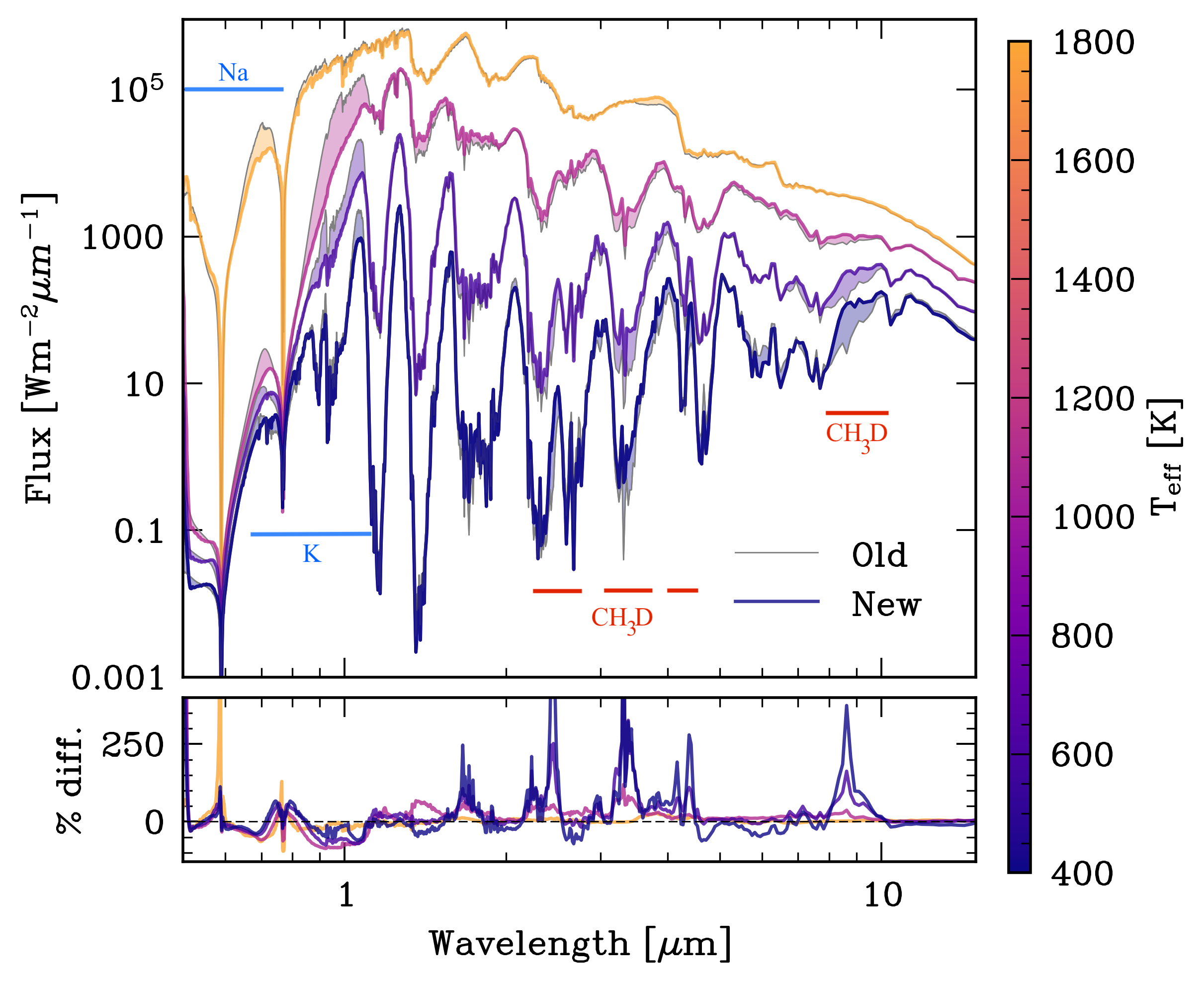}
    \caption{\textit{Top panel}: comparison of the new \exoII cloudless grid (coloured) to the previous grid (in grey) \citep{Charnay2018}, for models with $\logg = 4$\,dex and solar C/O and [M/H]. The difference between both models at the same \Teff is shaded in. The most notable changes are due to change in Na and K line profiles and the revised CH$_3$D abundance. \textit{Bottom panel}: relative residual flux $(\Fnew-\Fold)/\Fold \times 100$.}
    \vspace{-0.15cm}
    \label{fig: gridcomp}
\end{figure}

\paragraph[CH4]{CH$_{\mathsf{4}}$.} \exo already included its main isotopologue $^{12}$CH$_4$ as well as $^{13}$CH$_4$ and $^{12}$CH$_3$D. Unlike the other two, $^{12}$CH$_3$D is asymmetrical and has different vibration modes, with prominent band absorption at 2.7--2.9\,\mic, 3.2--3.4\,\mic, 4.25--4.66\,\mic, 6.5--6.9\,\mic, and 8.0--9.0\,\mic \citep{Wilmshurst1957, Rey2014}. This makes it detectable and separable in spectra, and we realised that \exo, despite citing a D/H ratio of $2\times 10^{-5}$ in \citet{Blain2021}, included a remarkable excess of $^{12}$CH$_3$D, leading to an over-absorption at CH$_3$D-absorbing wavelengths. This was especially relevant for spectra at the lower range of effective temperatures, where methane is more abundant since it is more stable at low temperatures \citep{Burrows1999}, as it is apparent in Fig.~\ref{fig:vmrs}. The unweighted methane isotopologue absorption cross sections, as well as the corrected, properly weighted (according to solar system abundances) isotopologues are shown in Fig.~\ref{fig: ch4_iso}. We therefore included the same isotopologues but with revised abundances compared to the most recent model. Fig.~\ref{fig: gridcomp} shows that this correction led to a weaker absorption in the $^{12}$CH$_3$D-absorbing ranges. 

\paragraph{Na and K.} Since the atmospheres of giant planets and BDs are cool enough to host neutral alkali atoms, their absorption lines play a major role in shaping the observed spectra \citep{Burrows2001, Burrows2002, Burrows2003}. The opacity in the visible and near-infrared range up to $\sim$1.2\,\mic is dominated by the heavily pressure-broadened wings of these alkali resonance lines. For K~{\sc i}, these lines are centred at  $\lambda = 0.766$ and 0.770\,$\mu$m (K D$_2$ and D$_1$ respectively), and for Na~{\sc i} they are centred at $\lambda = 0.589$ and 0.590\,$\mu$m (Na D$_2$ and D$_1$ respectively). These alkali doublets are produced by perturbations from molecular hydrogen and helium \citep{Allard2003}, so that they act as a pseudo-continuum. For most self-luminous gas giants and BDs, their flux peaks close to 1\,\mic; this emission originates from deep layers of the photosphere, where temperatures reach 1000\,K and pressures range between 10 and 100~bar. Thus, the line profile chosen for these alkalis is of utmost importance in this model as, for the \exo range of temperatures for cool BDs and hot giant planet atmospheres, they appreciably dampen the flux at its peak. It has become clear that, with the alkali line wings creating absorption up to 0.4\,\mic from line centre, the classical adoption of a Lorentzian profile is not appropriate in simulating their absorption in He and H$_2$-dominated atmospheres (see also \citealt{Allard2025}).

\begin{figure}[!t]
    \centering
    \includegraphics[width=\largimgunecol]{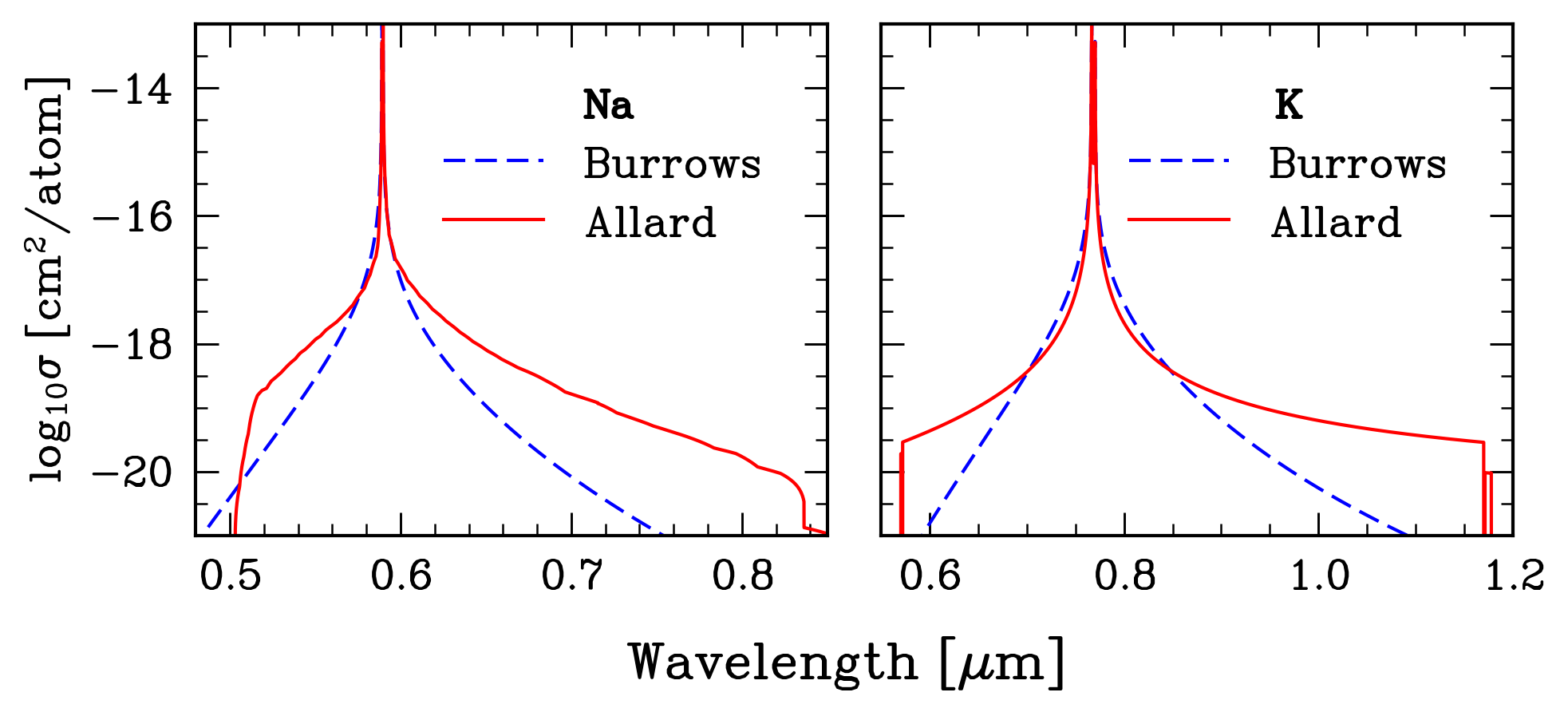}
    \caption{Comparison of alkali absorption cross sections using
    line lists of \citet{Allard2019} and \citet{Burrows2003} for Na (left panel), and \citet{Allard2016} and \citet{Burrows2003} for K (right panel), at $P = 10$\,bar, $T = 899.5$\,K, which are conditions in the atmosphere probed by the $\approx1$\,\mic peak flux.}
    \vspace{-0.15cm}
    \label{fig: na_k}
\end{figure}

New ab initio calculations of the theoretical potentials for the interactions of alkali atoms with He and H$_2$ were thus carried out in \citet{Allard2003, Allard2019}, and validated by laboratory measurements. This motivated our decision to incorporate in \exo the Na and K opacities calculated in \citet{Allard2016, Allard2019} respectively rather than those detailed in \citet{Burrows2003}, as had been done previously. Fig.~\ref{fig: na_k} shows the differences in the absorption cross section between the former and new adopted alkali profiles, with the \citet{Allard2016, Allard2019} profiles showing substantially more absorption at the wings. As seen in the resulting full simulated spectra in Fig \ref{fig: gridcomp}, the difference is non-negligible at 0.6--1.2\,\mic and results in a lower pseudo-continuum than previously computed for a range of effective temperatures.

\subsubsection[k-coefficients]{$k$-coefficients}
The chemical opacities described above are, in practice, represented in \exo in the form of correlated-$k$ distributions for each species. This approach is a way of binning down high resolution cross sections to evaluate radiative transfer at lower resolutions \citep{Lacis1991}. This approximation is more precise than simply binning down cross sections to lower resolutions, and more computationally efficient than evaluating the radiative transfer, line-by-line, at the native cross section resolution ($R=10^6$) and subsequently binning down, as demonstrated in \citet{Leconte2021}. For the low-resolution grid, the binning down was carried out with the Python library \exok, which distributes the cross section values, for each molecule, into $\Delta \nu$ bins of a constant 20 cm$^{-1}$ width, so that the resolution has a value of $R = 500$ at 1\,\mic. In each of these spectral intervals, the high-resolution cross sections were then sorted into 16 Gauss-Legendre quadrature points, organised in order of strength of absorption, where one radiative transfer calculation is carried out for each \citep{Baudino2015}. Similarly for the medium-resolution grid, the binning down of high-resolution cross section data was also carried out by \exok, but into bins of varying width so as to keep the spectral resolution constant ($R = 10,000$) throughout the wavelength range. At very high resolutions, the correlated-$k$ distribution approximation is no longer numerically faster than simply computing the radiative transfer at the native resolution: computing radiative transfer at $R = \textrm{70,000}$ for example would, with the correlated-$k$ method, require approximately 1,120,000 calculations (70,000 $\times$ 16), while simply evaluating line-by-line at $R=10^6$ would require 1,000,000 calculations and be more accurate \citep{Leconte2021}. In anticipation of the subsequent upgrade of \exoII to high resolutions of $R = \textrm{200,000}$ (Radcliffe et al.\ in prep.), we chose here to start from high resolution cross sections, allowing us to use the same data across grids of different resolutions, so as to render them completely consistent with each other.

\subsection{Cloud scheme}
Clouds of condensates represent a central source of absorption in Eq.~\ref{eq: extinction coefficient} when computing radiative transfer for BDs and planetary-mass companions. Clouds are governed by a balance between the downwards sedimentation of cloud particles due to gravitational settling versus the upwards turbulent mixing of condensate and vapour, as well as the evaporation of cloud particles and condensation back into clouds. We consider that, at equilibrium, the upwards mixing of condensates and vapour is balanced by the downwards flow of condensate caused by sedimentation \citep{Ackerman2001}:
\begin{equation}
    \frac{\partial q_c}{\partial z} = - \frac{\partial q_v}{\partial z} - \frac{\vsed}{\Kzz} q_c,
\label{eq: mass mixing ratios condensates}
\end{equation}
with $q_c$ and $q_v$ denoting the mass mixing ratios of condensates and vapour respectively, \Kzz being the eddy diffusion coefficient and \vsed the sedimentation velocity. Meanwhile, vapour condenses into clouds at supersaturation, when its pressure exceeds its saturation vapour pressure, $p_v > p_s$. The computation of mass mixing ratio of a condensate at a given level is computed by solving Eq.~\ref{eq: mass mixing ratios condensates}, using a formula for \Kzz based on mixing length theory \citep{Ackerman2001}, and finally the assumption that particles fall at their terminal velocity, all of which is detailed in \citet{Charnay2018}. 
In this model we make the assumption that supersaturation is weak (meaning condensation is efficient), i.e. $q_v \approx q_s$ above cloud level, $q_v$ and $q_s$ denoting the mass mixing ratio of vapour and that at saturation respectively. We included iron (Fe) and forsterite (Mg$_2$SiO$_4$) clouds, using pre-computed tables of optical properties (single scattering albedo, asymmetry factor, and extinction coefficient), assuming spherical particles following a log-normal size distribution with an effective variance of 0.3, for a range of wavelengths and mean particle radii \citep{Morley2012, Baudino2015, Blain2021}. As done previously in \citet{Charnay2018}, we included different ways of computing the cloud particle radii, either with a fixed sedimentation parameter or with simple microphysics. These different approaches are described in the following sections.

\subsubsection{Fixed sedimentation parameter}
\begin{figure}[!t]
    \centering
    \includegraphics[width=\largimgunecol]{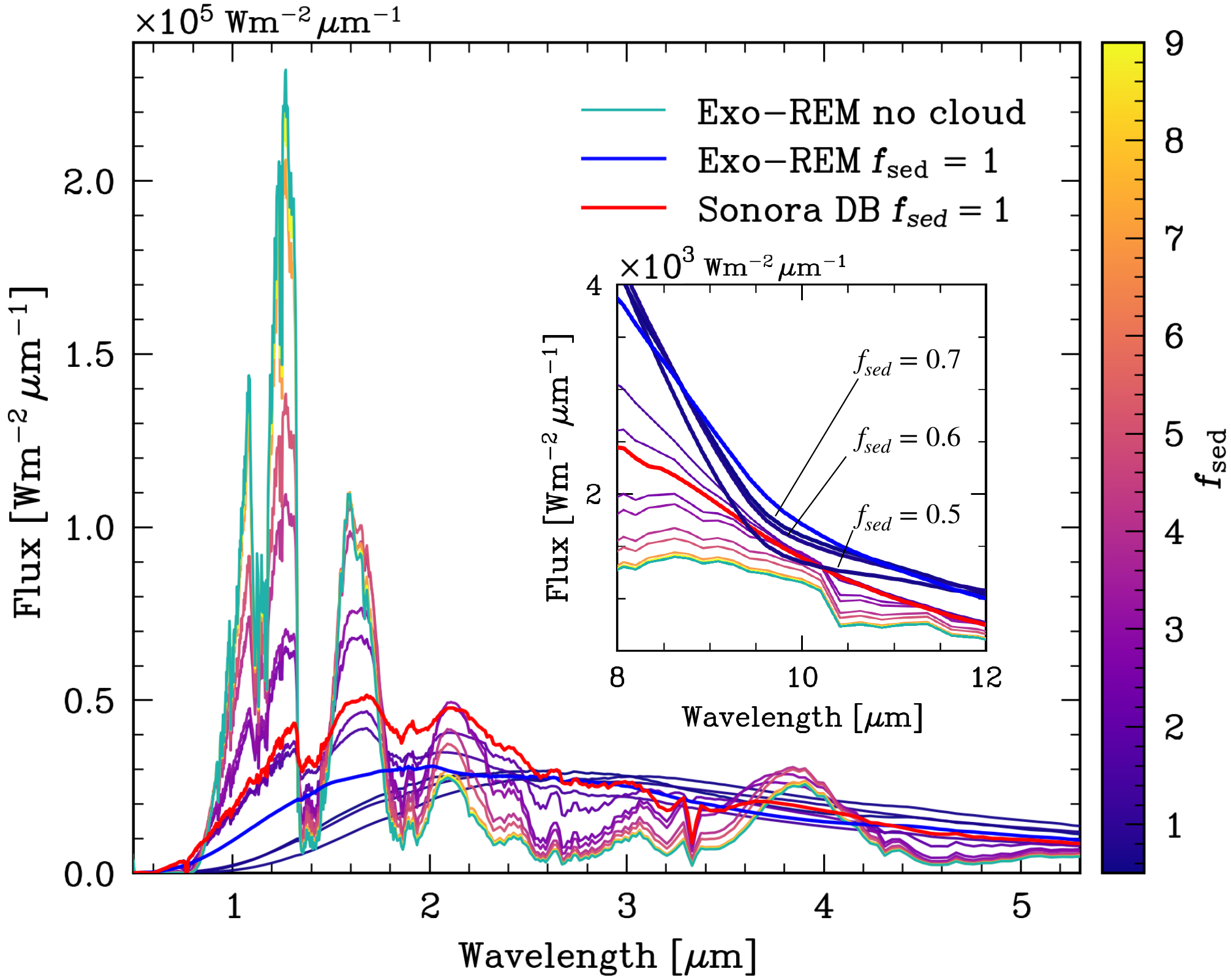}
    \caption{Spectra from the $R = 500$ \exoII \fsed grid; from almost cloudless (\fsed = 9) to extremely optically thick clouds (\fsed = 0.5). The 1--2\,\mic flux decreases with \fsed, with the thickest cloud cover dampening spectral features the most. In red is a \texttt{Sonora Diamondback} model with \fsed = 1, the lowest value explored in their grid; the corresponding \exoII model is shown in blue. The other parameters are fixed at $\Teff = 1200$\,K, $\logg = 4.5$\,dex, and solar C/O and [M/H]. The 8--10\,\mic inset highlights which \fsed values exhibit the silicate absorption.}
    \label{fig: fixed fsed}
\end{figure}

\citet{Ackerman2001} suggested that the vertical distribution of cloud particle radii could be simply represented by fixing the sedimentation parameter, which describes the ratio of sedimentation velocity to vertical mixing velocity, and is given by:
    \begin{equation}
        \fsed = \frac{\vsed H}{\Kzz}
    \end{equation}
with $H$ being the atmospheric scale height. Fixing this, in turn, fixes the mean particle radius as $\vsed \propto r^2$. Higher \fsed values lead to larger mean cloud particle radii, which sediment faster and render the cloud less optically thick. The \fsed value is held constant at each vertical level in the atmosphere, so that the ratio of characteristic upwards mixing timescale \taumix to downwards sedimentation timescale \taused is constant. Here, we chose to fix \fsed from a value of $\fsed=0.5$, where the clouds have almost no sedimentation, to $\fsed~=~9$, a value at which cloud particles sediment almost instantly (corresponding to an almost completely cloudless case). Large optical depths of the cloud layer significantly reduce the relatively transparent spectral windows, more noticeably in the 1-2 µm range, so that spectra with $\fsed~=~0.5$ resemble that of a blackbody, as seen in Fig.~\ref{fig: fixed fsed}.

\subsubsection{Simple microphysics: a timescale-based approach}
\label{subsection: simple microphysics}
Much like what was done by \citet{Rossow1978} and \citet{Cooper2003} for Earth, Mars, Venus, and Jupiter, in this approach we compare the timescales of the competing processes controlling cloud microphysics. At a given pressure, the process with the shortest timescale is said to dominate, driving particle size and thus dictating how quickly particles can grow and settle, while the other processes are neglected. The time scales associated with different processes are:
\begin{enumerate}[(a)]
\item vertical mixing ($\taumix = H^2/\Kzz $): the time it takes for particles to be vertically moved by atmospheric turbulence over a scale height;

\item sedimentation ($\taused = H/\vsed$): the time it takes for particles to fall through the scale height;

\item condensation growth (\taucond): the time required for a particle to grow by accreting surrounding vapor, described by the supersaturation parameter ($ = (q_v-q_s)/q_s$);

\item coalescence (\taucoal): the characteristic time for particles to collide and merge during sedimentation, a process that limits the mean radius of particles by quickly removing particles larger than a certain size.
\end{enumerate}
More details on this can be found in \citet{Charnay2018}.
\subsection{Numerical convergence}
\label{subsection: numerical convergence}

In general, to run the \exo model, an initial trial pressure--temperature profile ($P$--$T$) describing the thermal structure of the atmosphere must be input, then the algorithm searches for a $P$--$T$ profile that ensures conservation of flux at each pressure level. The model computes the chemistry iteratively until it converges to a radiative equilibrium solution, or until it has reached the maximum number of iterations. It is therefore necessary to input a $P$--$T$ profile that is as close to the final profile as possible to minimise the number of iterations and maximise the likelihood of convergence. However, in practice, convergence does not always occur. We outline in Section~\ref{subsection: robust convergence criteria} the three markers for an ``unconverged'' spectrum, and their treatment, allowing us to impose benchmark convergence criteria, as a fail-safe, to the \exo model. \exoII has been completely cleaned of these, leaving gaps where convergence was not achievable.

\subsection{Resolution}
The \exoII low-resolution grids have a spectral resolution of $R = 500$ at 1\,\mic. However, the step size is constant in wavenumber, with $\Delta \nu = 20$~cm$^{-1}$, meaning that the spectral resolution is not constant across wavelengths (since $R = \lambda/\Delta \lambda$), reaching values of 50 at 10\,\mic and as low as 2 at 250\,\mic. Furthermore, the binning step is also $20$\,cm$^{-1}$, meaning that there is one point per spectral element and the binning resolution, $R_b$, is equal to the spectral resolution; a correct binning should be at most 1/2 of the width of the spectral bins to ensure Nyquist sampling. Therefore, the spectral resolution should be downgraded (by convolution) to $R = 250$ at 1\,\mic and so on.

Meanwhile, the \exoII medium-resolution spectra have a true spectral resolution of 10,000: they were made by recomputing the radiative transfer at a resolution of $R = \textrm{30,000}$ from the $R = 500$ chemical profiles, then subsequently downgrading the resulting spectra to a final, constant spectral resolution of $R = \textrm{10,000}$, so as to provide a super-Nyquist sampling at $R_b = \textrm{30,000}$. This was done using \exok \citep{Leconte2021}, and the same $R = 10^6$ cross section files were used as for making the $R = 500$ $k$-tables, but this time binned down to $R = \textrm{30,000}$ $k$-tables. The spectra in this medium-resolution grid, when binned down to the same resolution, are virtually identical to the corresponding $R = 500$ spectra, with differences in flux arising from discrepancies when evaluating radiative transfer at different resolutions.

\section{Results and applications}
\label{section: results}

\subsection[Exo-REM k26: the new generation of models]{\exoII: the new generation of models}
\label{subsection: comparison between old and new models}
To summarise, we therefore produced three low-resolution grids of models: one cloud-free, one with simple microphysics and lastly one with an \fsed dimension spanning $0.5\le\fsed\le9$ , the range of which is shown in Fig.~\ref{fig: fixed fsed}. We additionally produced one new medium-resolution grid with the extra \fsed dimension at a constant resolution of $R = \textrm{10,000}$. All of these are publicly available \footnote{\url{https://lesia.obspm.fr/exorem/YGP_grids/Exo-REMk26/}}.

\subsection[Reproducing the L-T transition]{Reproducing the L--T transition}
\label{subsection: reproducing the l--t transition}

\begin{figure}[!t]
    \centering
    \includegraphics[width=\largimgunecol]{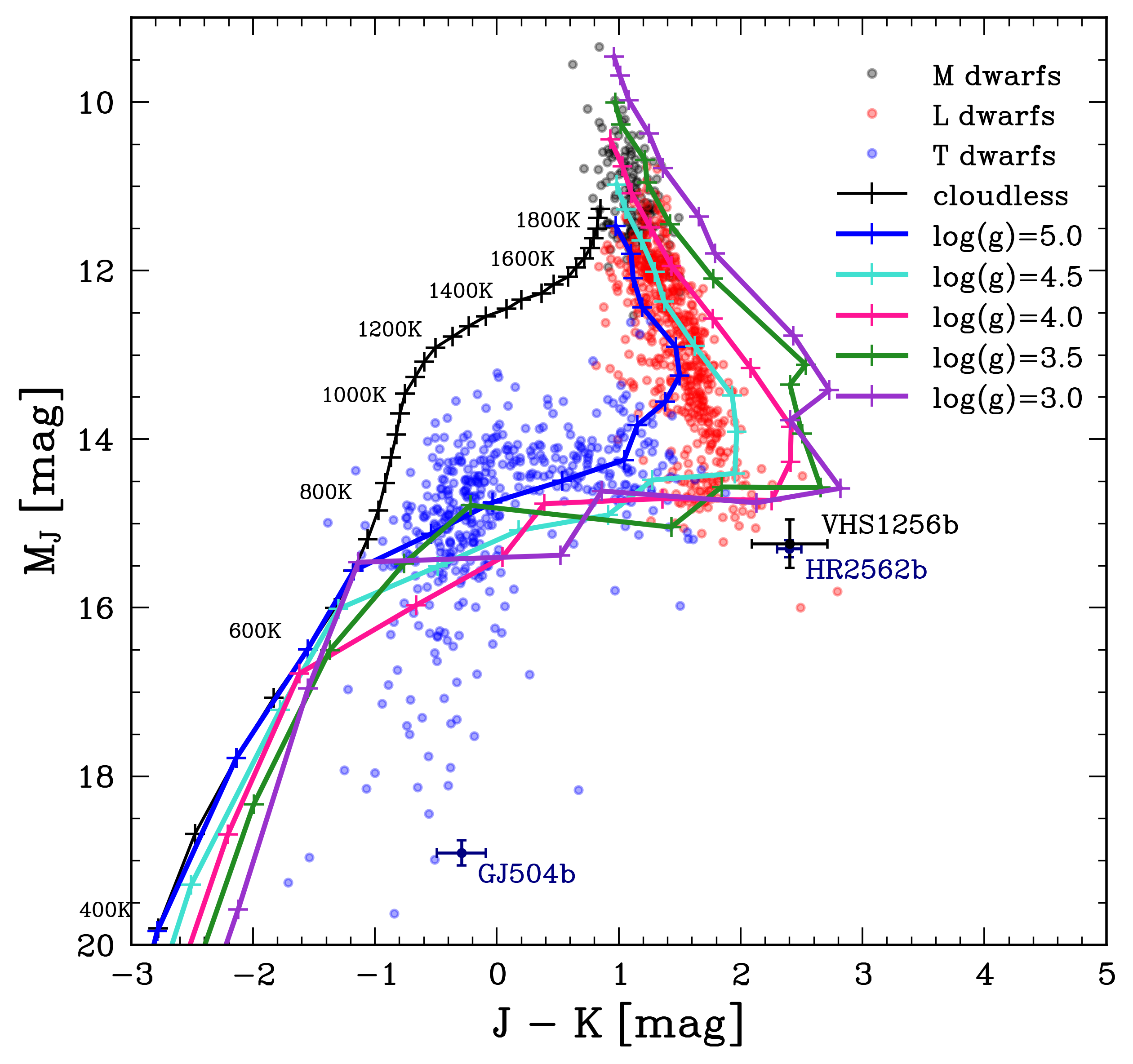}
    \caption{CMD with the computed $J$ vs. $J-K$ magnitudes for the \exoII cloudless $\logg = 5$ (in black) and simple cloud microphysics models (for $\logg = 3$, 3.5, 4, 4.5, and~5 in purple, green, pink, light blue, and blue respectively), at constant solar C/O and [M/H]. The simple microphysics grid was computed with a supersaturation parameter $s = 0.1$. The M, L, and T dwarfs are plotted in black, red and blue dots respectively \citep{Best2025}. Data for GJ~504\,b, \vhs and, HR~2562\,b are from \citet{Kuzuhara2013}, \citet{Gauza2015}, and \citet{Konopacky2016}, respectively.}
    \label{fig:ltmicrophysics}
\end{figure}

\begin{figure}[!t]
    \centering
    \includegraphics[width=\largimgunecol]{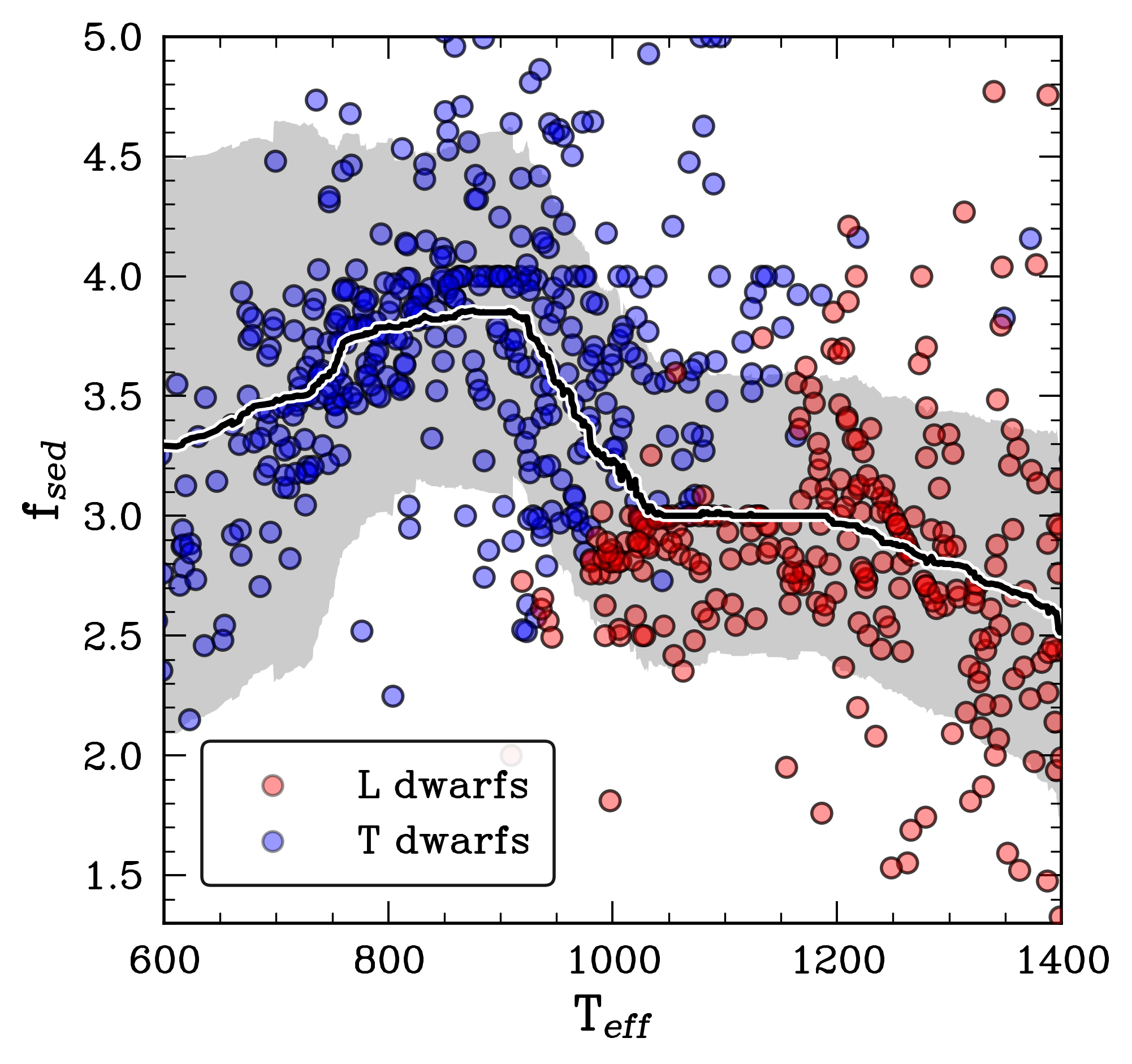}
    \caption{Evolution of \fsed for BDs over the L--T transition, found by computing the magnitudes of \exoII spectra of varying \fsed and \Teff values at $\logg = 4.5$ and solar C/O and [M/H], then interpolating to find the \fsed and \Teff at the \citet{Best2025} data points. The median over 20\,K \Teff ranges is shown in black, and the 1$\sigma$ region is shaded in grey.}
    \label{fig:lt_fsed}
\end{figure}

The $J-K$ (MKO) magnitudes of spectra in the cloudless and simple microphysics grids are shown on a CMD in Fig.~\ref{fig:ltmicrophysics}: as previously, the cloudless models do not follow the L--T transition, but instead keep steadily decreasing in $J$ magnitude and increasing in $J-K$ as temperature increases. Conversely, the simple microphysics models approximately follow the shift to bluer colours around $\Teff = 1400$\,K and cooler, as the forsterite and iron clouds appear and then dissipate at lower effective temperatures. While the simple microphysics grid provides a satisfactory cloud description, the grid with the added \fsed dimension ensures that each and every object on the CMD is represented, even the reddest ones such as \vhs. Computing the $J-K$ magnitudes of our model spectra for varying \fsed and \Teff, and interpolating onto observed L and T dwarf magnitudes (see Fig.~\ref{fig:lt_fsed2}) results in a general trend along the L--T transition, whereby \fsed values increase (from $\fsed \sim 2.7$ to $\sim 3.9$), as shown in Fig.~\ref{fig:lt_fsed}. This trend was predicted in \citet{Saumon2008} and has been further substantiated with the X-SHYNE (X-SHooter medium-resolution near-infrared survey for Young, Nearby Exoplanet analogs) survey of 43 spectra of isolated BDs: \citet{Petrus2025} found, with \texttt{Sonora Diamondback}, a clear trend of increasing \fsed from L1 to T7 spectral types.

\subsection[GJ 504 b: impact of the CH3D correction]{GJ~504\,b: impact of the $\text{CH}_3\text{D}$ correction}
\label{subsubsection:gj504b}
To exemplify the change in results brought about by the methane D/H correction in \exo detailed in Section~\ref{subsubsection: line lists}, we chose the case of GJ~504\,b, a very cool object that exhibits marked methane absorption \citep{Janson2013}, and has been analysed with the previous \exo grid in \citet{Malin2025}. 
\begin{figure*}[t]
    \includegraphics[scale = 0.59]{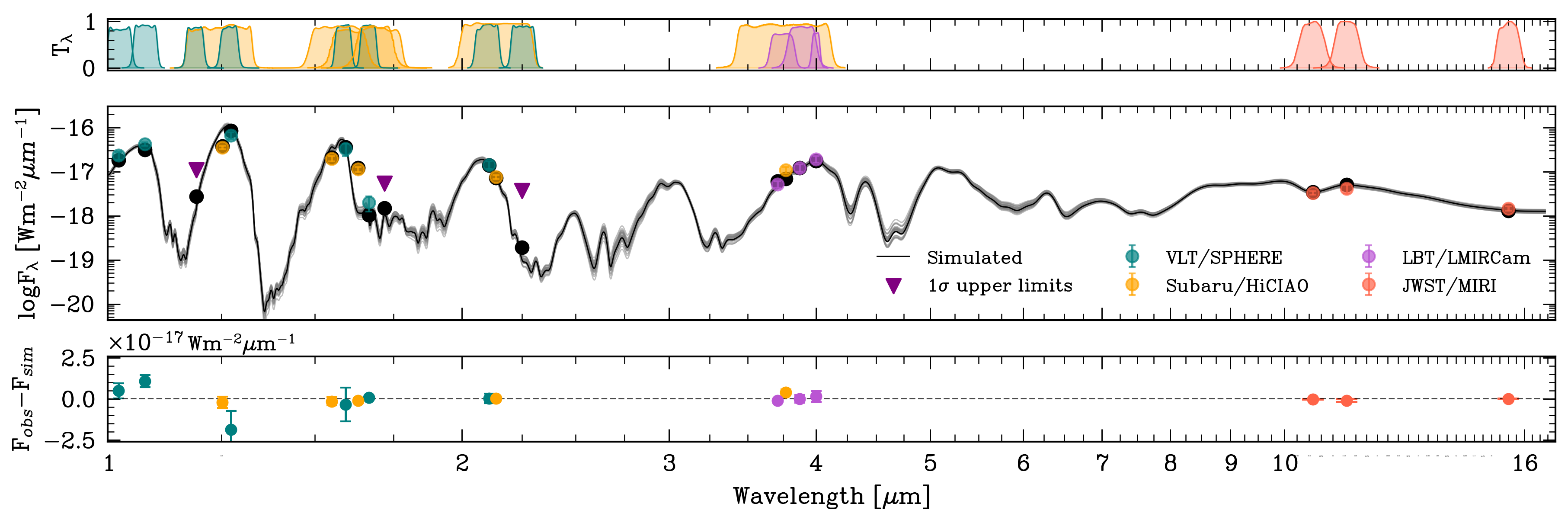}
    \caption{Best fit for GJ~504\,b photometry with revised parameters found with \exoII, posterior distributions of which are provided in Fig.~\ref{fig: gj504b corner}. Plotted in grey are random samples from the family of spectra with a likelihood within 1 $\sigma$ of the best fit. \textit{Top panel}: normalised transmission for each filter. \textit{Bottom panel}: residual flux between the observed and simulated photometry.}
    \label{fig: bf_gj504b}
\end{figure*}
This faint companion was detected via direct imaging in 2011, in H-band ($\sim$1.6 $\upmu$m) Subaru/HiCIAO (the High-Contrast Instrument with Adaptive Optics) observations \citep{Kuzuhara2013}, at a wide orbit with a projected distance of 43.5\,au around GJ~504\,A. The primary had been observed in the course of the SEEDS (Strategic Explorations of Exoplanets and Disks with Subaru) survey at a distance of $17.56 \pm 0.08$\,pc, and reported to be a Sun-like main sequence star of spectral type G, with a mass of $M_{\star} = 1.2~\MSun$ and rather young age of 160$^{+350}_{-60}$\,Myr based on stellar gyrochronological and chromospheric activities \citep{Kuzuhara2013}. However, this was debated by \citet{Fuhrmann2015} and \citet{DOrazi2017}, who suggested an older stellar age, ranging from 1.5 to 4\,Gyr. They proposed that the high levels of rotation and chromospheric activity, typically indicative of a young stellar age, could result from the recent engulfment of a short-period hot Jupiter, which supported the argument for an older system. Finally, the most recent isochronal studies suggest that the star may either lie above the main sequence on the Hertzsprung--Russell Diagram and thus have an age of $4.0\pm1.8$\,Gyr, or be much younger with an age of $21\pm2$\,Myr \citep{Bonnefoy2018}.

Naturally, this degeneracy in age transfers to its companion GJ~504\,b, and gives rise to two distinctly different values for its mass. For the observed flux, it would have to be a relatively young gas giant or a much older BD: using the hot-start model from \citet{Baraffe2002} with an age of 160$^{+350}_{-60}$\,Myr \citep{Kuzuhara2013} leads to a mass of $M = 4.5^{+4.5}_{-1.0}\,\MJup$, while taking the isochronal ages of $21\pm2$\,Myr and $4.0\pm1.8$\,Gyr from \citet{Bonnefoy2018} would yield mass values of $M = 1.3^{+0.6}_{-0.3}$\,\MJup or $M = 23.3^{+10}_{-9}$\,\MJup respectively. GJ~504\,b exhibits notable differences compared to previously observed exoplanets, displaying a bluer colour ($J - H = -0.23$\,mag) \citep{Kuzuhara2013} and being one of the coolest companions ever observed directly, with an effective temperature of $\sim 500$\,K \citep{Malin2025}. This, along with its discrepancy in age and mass, is a powerful incentive for further investigation of GJ~504\,b and how it formed. 

We conducted forward modelling with the low-resolution \texttt{Exo-REM k26} grid using \texttt{ForMoSA}\footnote{\url{https://formosa.readthedocs.io}} \citep{Petrus2023}, a Bayesian forward-modelling tool that uses a nested sampling algorithm for parameter exploration with a given likelihood, as done for example in \citet{Petrus2021,Petrus2023} or \citet{Palma-Bifani2023,Palma-Bifani2024}. We used 4,000 live points and uniform priors spanning the whole parameter space. We included the J2, K2, and CH$_4$L photometry 3$\sigma$ upper limits of by adding a component to the total $\chi ^2$ in the likelihood function: considering a non-detection of flux $\FUL = \mubruit + 3\sigUL$ with an average, $\mubruit = 0$, and dispersion \sigUL for each upper limit filter, we used a cumulative distribution function to contribute to the total $\chi^2$ when the model flux approaches the upper limit from below, or lies anywhere above, similarly to \citet{Sawicki2012}. The total goodness-of-fit is
\begin{equation}
\chi^2_{\rm tot} =
\sum_{i\,\in\,{\rm det}}\frac{\left(F_{i,\,{\rm obs}}-F_{i,\,{\rm mod}}\right)^2}{\sigma_i^2}\notag
-\sum_{j\,\in\,{\rm UL}}2\,\ln 
\Phi\!\left(\frac{F_{j,\,{\rm UL}} - F_{j,\,{\rm mod}}}{\sigma_{j,\,{\rm UL}}}
\right),
\end{equation}
where $\Phi(\cdot)$ denotes the standard normal cumulative distribution function. Fig.~\ref{fig: gj504b corner} shows the resulting posterior distributions. We obtained an effective temperature of $\Teff = 473^{+14}_{-12}$\,K, a surface gravity of $\logg = 4.0 \pm 0.1$, a radius of $R = 1.16^{+0.08}_{-0.07}$\,\RJup, a metallicity of $[M/H] = 0.9\pm 0.1$\,dex, a carbon to oxygen of $\co = 0.71^{+0.05}_{-0.06}$, and \fsed of $2.6^{+0.2}_{-0.3}$, as shown in Fig.~\ref{fig: bf_gj504b} and summarised in Table. \ref{tab:exorem_summary}, corresponding to a \chisqreduced~=~1.38. This yielded an inferred luminosity of $\log L/\LSun = -6.19 \pm 0.02$ and mass of $M = 5.4 ^{+1.5}_{-1.4}$\,\MJup respectively. The mass that we found is in accordance with the age of 160$^{+350}_{-60}$\,Myr from \citet{Kuzuhara2013}. However, there are biases in purely atmospheric forward modelling analyses for mass determination; we have not coupled the model with an interior or thermal evolution model, so we cannot completely exclude the other possible ages (and thus masses).

\begin{table}[t]
\caption{Best fit parameter values for the GJ~504\,b photometry using the former version of \exo vs. using \exoII.}
\label{tab:exorem_summary}
\centering
\small
\renewcommand{\arraystretch}{1.3}
\setlength{\tabcolsep}{3pt}
\begin{tabular}{cccccccl}
\hline\hline
\begin{tabular}[c]{@{}l@{}}\Teff\\ {[}K{]}\end{tabular} &
\begin{tabular}[c]{@{}l@{}}\logg\\ {[}dex{]}\end{tabular} &
\begin{tabular}[c]{@{}l@{}}{[}M/H{]}\\ {[}dex{]}\end{tabular} &
C/O & \fsed &
\begin{tabular}[c]{@{}l@{}}$R$\\ {[}\RJup{]}\end{tabular} &
Study \\
\hline
$512\pm10$ & 
$3.45^{+0.35}_{-0.25}$ &
$0.54^{+0.09}_{-0.11}$ &
$0.70^{+0.06}_{-0.07}$ &
$\cdots$ &
$1.08^{+0.04}_{-0.03}$ & 
(1) \\
$473^{+14}_{-12}$ &
$4.0\pm0.1$ &  
$0.9\pm0.1$ & 
$0.71^{+0.05}_{-0.06}$ &
$2.6^{+0.2}_{-0.3}$ &
$1.16^{+0.08}_{-0.07}$ & 
   \\
(2)   \\
\hline
\end{tabular}
\tablefoot{
(1)~\citet{Malin2025}, using the older version of \exo.
(2)~This work, using \exoII.
}
\end{table}

\subsection[VHS 1256 b: a temperamental companion]{\vhs: a temperamental companion}
\label{subsection: vhs 1256b}
Displaying strong observational pointers towards thick, patchy silicate cloud cover, the elusive \vhs presents an excellent case to stress-test the medium-resolution \exoII models, with their wide range of \fsed values. It was discovered using the VISTA Hemisphere Survey \citep{Gauza2015} around the M dwarf binary VHS J125601.92--125723.9 at a projected separation of 105\,au \citep{Gauza2015, Rich2016, Stone2016}. 

Originally, the host star was thought to be a single M dwarf (and not yet known to be a binary), and a distance measurement of $d=12.7\pm1.0$\,pc was initially determined by \citet{Gauza2015}, which suggested a planetary mass of $M=11^{+10}_{-2}$\,\MJup for the companion. However, once the star was found to be an equal-mass binary, the original distance measurement severely conflicted with the expected luminosities of the host star components, and the spectrophotometric distance of $d=17.2 \pm 2.6$\,pc was found by \citet{Stone2016}. This was definitively resolved by high-precision astrometry from \citet{Dupuy2020} with the Canada-France-Hawaii Telescope, which yielded a revised distance measurement of $d=22.2^{+1.1}_{-1.2}$\,pc. This dramatically increased the expected intrinsic luminosity and thus -- through the \citet{Saumon2008} evolution model -- mass, effective temperature, and surface gravity to $M=19\pm5$\,\MJup, $1240\pm50$\,K, and 4.55$^{+0.15}_{-0.11}$\,dex respectively, placing it in the BD mass realm. Recently, \cite{Dupuy2023} derived an age of $140\pm20$\,Myr for the system using the \citet{Baraffe2015} evolutionary model; with \vhs's observed luminosity, this places it in a region of overlap between deuterium-inert and deuterium-fusing evolutionary model tracks from \citet{Saumon2008}. This thus yields two possible masses of $M = 12.0 \pm0.1$\,\MJup and~$16\pm1$\,\MJup, corresponding to a deuterium-bearing or deuterium-depleted object respectively. The lower-mass scenario would have a corresponding $R = 1.3\RJup$, \Teff $= 1153 \pm 5$, and \logg $ = 4.268 \pm 0.006$, while the higher-mass scenario would correspond to $R = 1.22\RJup$, \Teff $= 1194 \pm 9$, and \logg $ = 4.45 \pm 0.03$ \citep{Dupuy2023}. Through light curve fitting from 36-h Spitzer/IRAC monitoring, this companion has been attributed a rotational period of $P=22.04 \pm 0.05$\,h \citep{Zhou2020}, the upper limit of BD rotation periods. \citet{Poon2024} found a line-of-sight spin axis inclination of $i_p = 90 \pm 18^{\circ}$, meaning that the viewing geometry is close to edge-on.

\vhs exhibits clear signs of vigorous atmospheric dynamics. It is the most variable object observed to date, shows a shallow $L$-band methane absorption indicative of disequilibrium chemistry \citep{Miles2018}, has an extremely red $J-K_s$ colour of 2.47\,mag \citep{Gauza2015} (see Fig.~\ref{fig:ltmicrophysics}), and displays pronounced silicate absorption -- together, these features point to intense vertical mixing and an optically thick, patchy photospheric cloud cover \citep{Madhusudhan2011, Marley2012}. As evidenced by \citet{Petrus2024} and \citet{Lueber2024}, the existing 1D pre-computed models have difficulties in recreating the multifaceted structure and dynamics of \vhs's atmosphere, and most strikingly do not reproduce the silicate absorption. To show this particular spectral feature, models necessitate optically thick cloud cover, a condition that can be produced by a low \fsed regime. This constitutes the motivation for an application of our new models, that do indeed explore these low values, on \vhs.

\subsubsection{Classical approach}
\label{subsubsection: classical approach}
\begin{figure}[t]
    \centering
    \includegraphics[width=\largimgunecol]{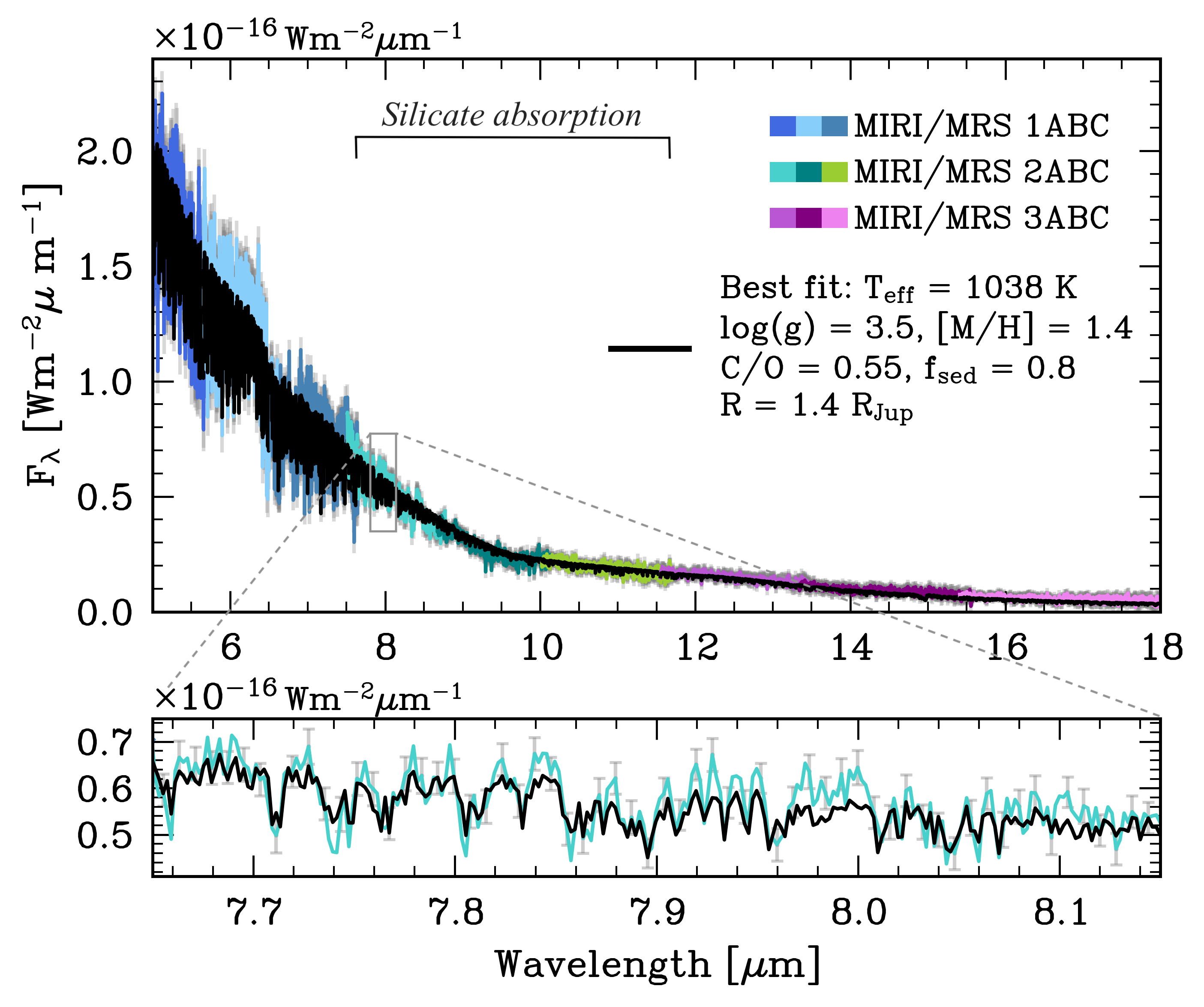}
    \caption{\textit{Top panel}: 1 column best fit in black for only the \vhs MIRI/MRS data (\citealp{Gauza2015}) from forward modelling with the medium resolution \exoII grid. Observation data are coloured according to MIRI channels. \textit{Bottom panel}: a zoom of the best fit. The $1\sigma$ uncertainty is shown in grey in the top panel, and, in the bottom panel, as grey error bars for one out of four points.}
    \label{fig: vhs best fit MIRI}
\end{figure}
\begin{figure*}[t]
    \centering
    \includegraphics[width=\largimgdeuxcols]{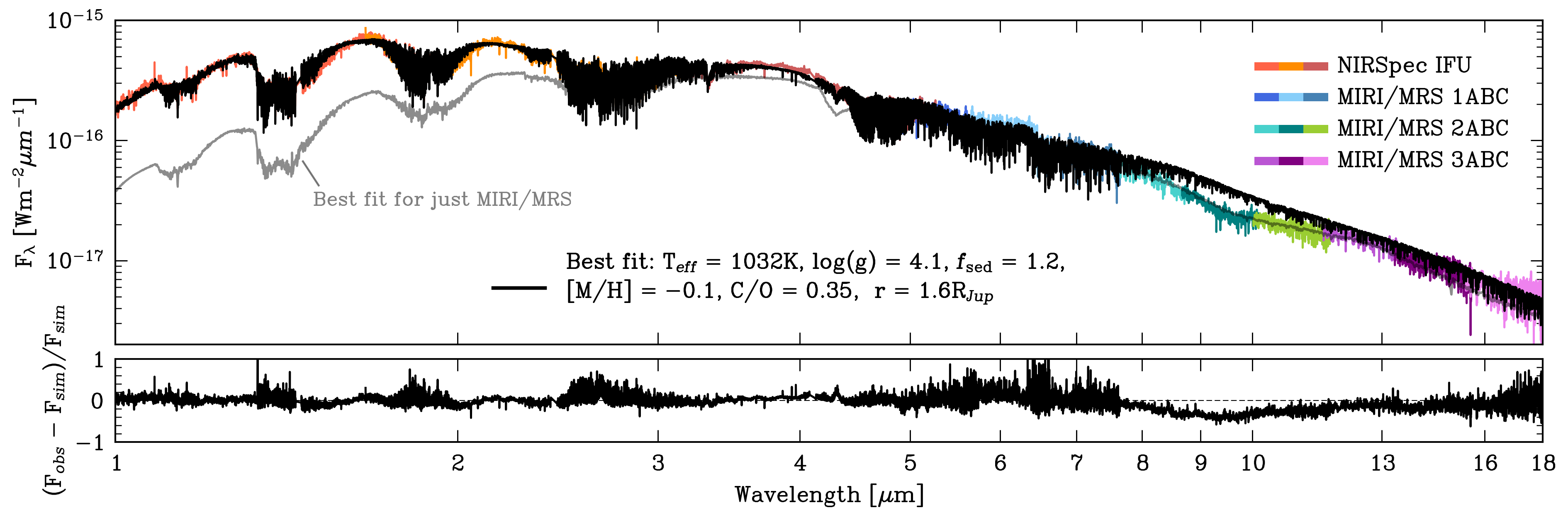}
    \caption{\textit{Top panel}: same as Fig.~\ref{fig: vhs best fit MIRI} but with the MIRI + NIRSpec \vhs observations (\citealp{Gauza2015}). The best fit for the MIRI/MRS spectrum only from Fig.~\ref{fig: vhs best fit MIRI} is shown in grey, with reduced resolution for visibility. \textit{Bottom panel}: the residual flux divided by the simulated flux.}
    \label{fig: vhs best fit 1column}
\end{figure*}
We performed forward modelling using \texttt{ForMoSA} taking the medium resolution \exoII models in tandem with the MIRI/MRS spectrum from \cite{Gauza2015}, setting 500 living points with flat priors exploring the entire parameter space. The corresponding best fit and posterior distributions are shown in Fig.~\ref{fig: vhs best fit MIRI} and Fig.~\ref{fig: vhs mrs corner} respectively; in particular we were able to reproduce the silicate absorption with, as expected, a low \fsed value of 0.82, as well as a super-solar metallicity and a \Teff of $1038$\,K, a temperature lower than previous estimates. With $\chisqreduced=1.96$, the overall shape of the pseudo-continuum is a good fit and the absorption features match up, but do not completely reflect the depth that is seen in the observations. 

To test a broader range of wavelengths, we ran \texttt{ForMoSA} in the same conditions, but also included the JWST/NIRSpec portion of the spectrum, so that the fitting spanned from 1 to 18\,\mic. The best fit model for NIRSpec + MIRI/MRS data is shown in Fig.~\ref{fig: vhs best fit 1column}, with posterior distributions in Fig.~\ref{fig: vhs all obs corner} corresponding to \Teff = 1032K, a much lower metallicity, and an overall \fsed = 1.2. This model does not show a pronounced silicate dip, but does nevertheless fit closer, with $\chisqreduced = 152$, than any attempted fitting with the previous \exo or current \texttt{ATMO}, \texttt{BT-Settl}, \texttt{DRIFT-PHOENIX}, and \texttt{SONORA Diamondback} model grids, which had $\chisqreduced = 246$, 334, 673, 301, and 522 respectively \citep{Petrus2024}. It should be noted that the best fit model previously obtained for only the MIRI data, shown in grey in Fig.~\ref{fig: vhs best fit 1column},  clearly fails to fit the lower wavelengths of the NIRSpec data. We found that the silicate feature could only begin to be accessed with $\fsed < 1$, but that the general disk averaged shape of the VHS~1256\,b JWST spectrum resembled one of $\fsed > 1$. This underscores the limitations of 1D models: the variability monitoring of \vhs strongly indicates heterogeneous cloud cover, which cannot be adequately described by a single 1D spectrum, with its limiting assumption of atmospheric homogeneity.

\subsubsection{A two-column approach}
\label{subsubsection: two column approach}

\begin{figure*}[!t]
    \centering
    \includegraphics[width=\largimgdeuxcols]{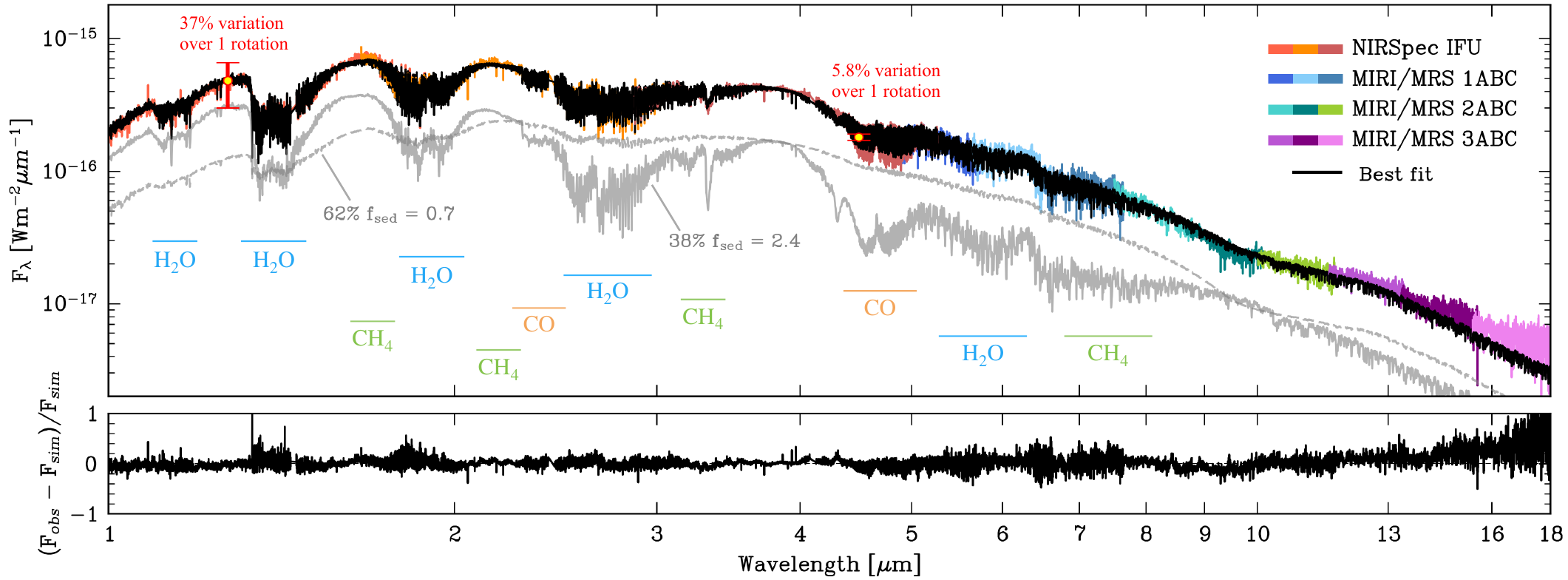}
    \caption{Final best fit for VHS 1256 b. \textit{Top panel}: same as Fig.~\ref{fig: vhs best fit 1column}, but using the two-column approach. Best fit parameters are $\Teff = 1153$\,K, $\logg = 4.0$\,dex, [M/H] = 0.08\,dex, C/O = 0.60, $R = 1.2\,\RJup$, with an $\alpha = 0.62$ so a linear combination of $\fsed = 0.7$ (62\%) in grey dashed and $\fsed = 2.4$ (38\%) in grey solid line. Both grey \fsed components have been scaled by their proportion and multiplied by a factor of 0.6 for visibility. The two \fsed component spectra are shown with reduced resolution for visibility. \textit{Bottom panel}: same as Fig.~\ref{fig: vhs best fit 1column}.}
    \label{fig: vhs best fit}
\end{figure*}

\begin{figure}[!t]
    \centering
    \includegraphics[width=\largimgunecol]{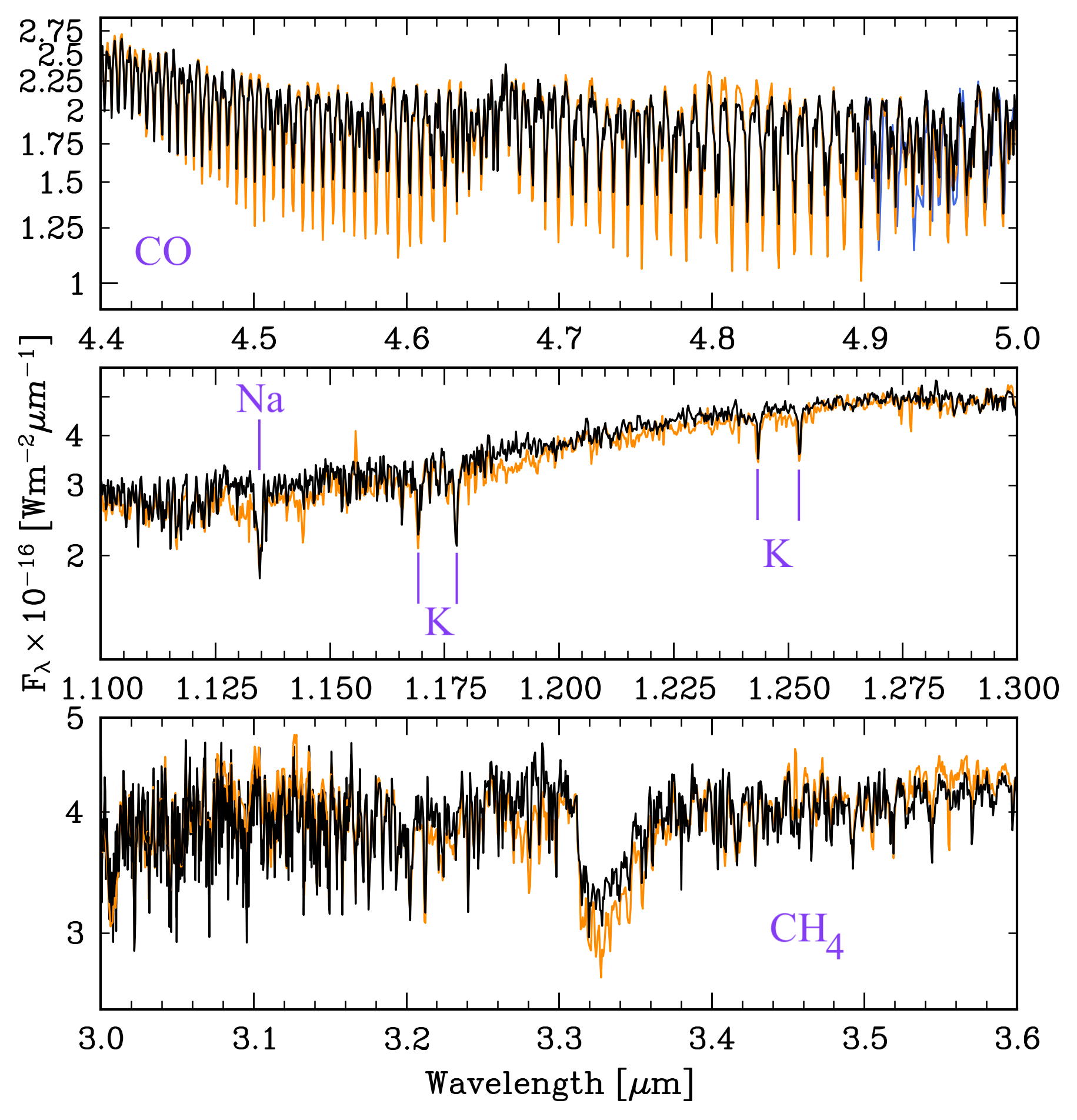}
    \caption{Zooms of the final two-column best fit model (black) and observation (orange) of \vhs showing, from top to bottom, primarily the CO, alkali, and CH$_4$ absorptions.}
    \label{fig: mol zooms}
\end{figure}

To attempt a non-homogeneous description of \vhs's atmosphere, we suggested that multiple \exoII models of different \fsed values would together imitate a patchy cloud cover scenario, in a ``1D+1D'' manner, similarly to \citet{Vos2023}, \citet{Zhang2025}, \citet{Mollire2025}, \citet{Marley2010}, and \citet{Morley2014b, Morley2014a}, the differences being discussed in Section~\ref{subsection: advantages and limitations of the two column approach}. A two-column approach was proposed for \vhs in \citet{Miles2023} using \texttt{PICASSO 3.0} \citep{Mukherjee2023}, precomputing an $\fsed=0.6$ and $\fsed=1.0$ model separately, then manually combining them linearly with a 90--10\% split, but the authors were not able to produce a satisfactory fit or reproduce the silicate cloud feature. We applied a similar approach but, by contrast, in a forward modelling framework, with pre-computed grids of self-consistent models, where each iteration explores a linear combination of two different models with common \Teff, C/O, [M/H], and \logg, allowing only \fsed to vary between both spectra. An $\alpha$ parameter (0$\leq \alpha \leq$1) was added to allow proportions of each \fsed model to vary, and thus alter the percentage of thick cloud coverage:
\begin{equation}
 \label{eq:Fcombo}
 \Fcomb = \alpha \Ffsedun + (1-\alpha)\Ffseddeux,
\end{equation}
each of the two columns with its own self-consistent thermal profile at radiative-convective equilibrium. We re-ran \formosa in these conditions with the whole JWST spectrum, which took $\sim88$\,h to converge on seven processors. The resulting posterior distributions are shown in Fig.~\ref{fig: 2fsed allobs vhs corner}. The condition $\fsedun < \fseddeux$ was imposed to avoid two identical solutions. The resulting best fit ($\chisqreduced= 84$) yielded  a common $\Teff = 1153$\,K, $\logg = 4.0$\,dex, [M/H] = 0.08\,dex, C/O = 0.60, and $R = 1.2\,\RJup$, with an $\alpha = 0.62$ for $\fsedun=0.7$ and $\fseddeux=2.4$, and is shown in Figs.~\ref{fig: vhs best fit} and~\ref{fig: mol zooms}. A summary of the retrieved best-fit parameters is shown in Table~\ref{tab:retrieval_results}, and a discussion on the misrepresentative nature of the narrow error bars on the posterior distributions found when forward modelling with the \vhs JWST data can be found in Section~\ref{subsection: posterior uncertainties}. Our retrieved \Teff and radius are consistent with the values found in \cite{Dupuy2023} for their lower-mass solution ($12.0\pm 0.1$\,\MJup). The mass value obtained from our retrieved \logg and radius ($M = 5.8\,\MJup$) differs from this, but given the uncertainty associated to mass retrieval by both atmospheric models (primarily related to gravity determination, see \citet{Petrus2023}) and evolutionary models (arising from assumptions regarding clouds, thermal structure, core composition, and initial conditions), our value remains reasonably in accordance with \citet{Dupuy2023}. Together, these mass estimates point towards \vhs being a planetary-mass object.

\begin{figure}[!t]
    \centering
    \includegraphics[width=\largimgunecol]{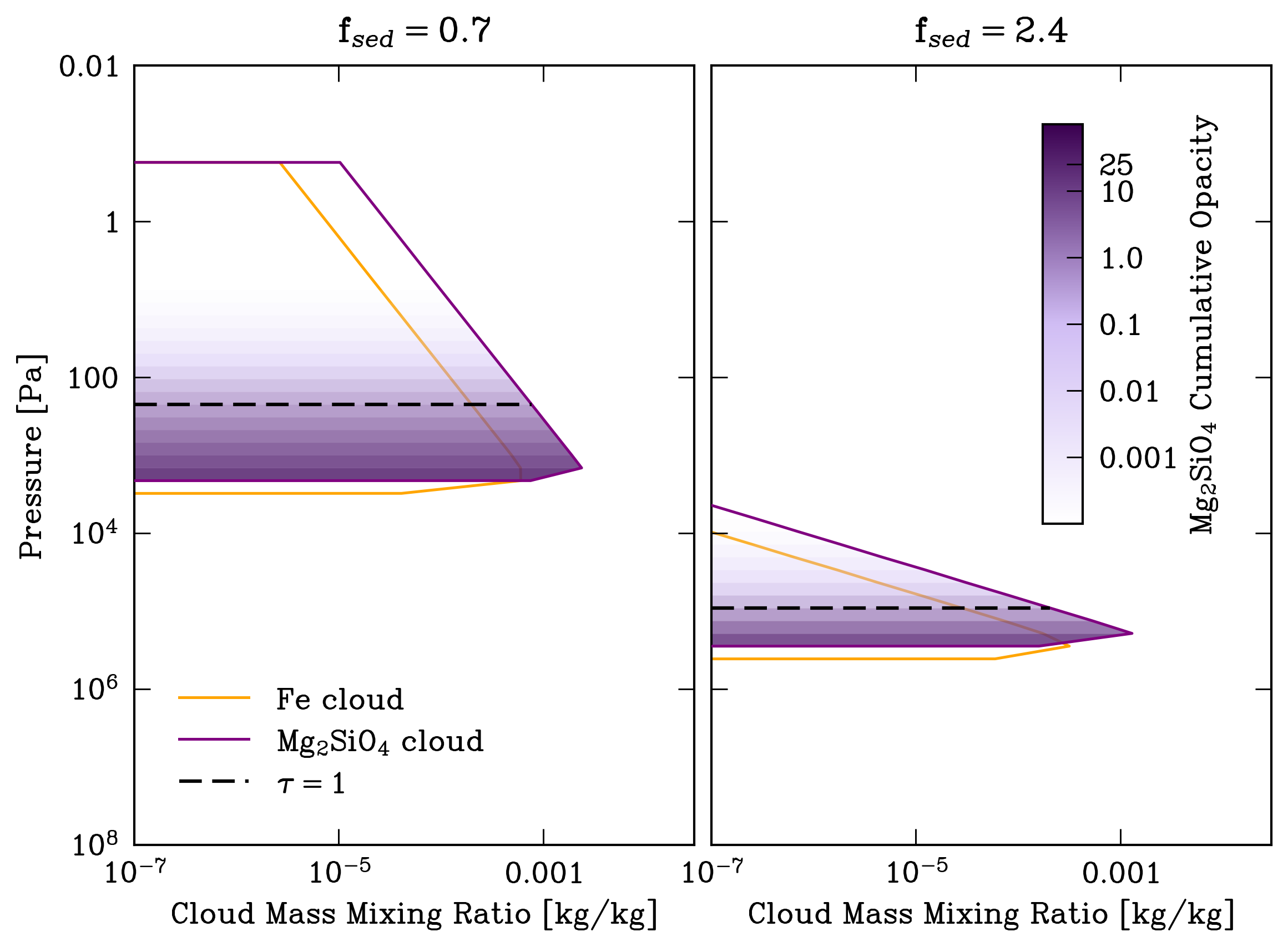}
    \caption{Vertical cloud structure for both \fsed columns used to describe the patchy cloud cover of \vhs. We show the cumulative forsterite (Mg$_2$SiO$_4$) cloud opacity (from the top layer) at $\lambda = 1$\,\mic and the pressure level corresponding to a cloud opacity of $\tau = 1$ (dashed).}
    \label{fig: mmrs}
\end{figure}

\begin{figure*}[!t]
    \centering
    \includegraphics[width=\largimgdeuxcols]{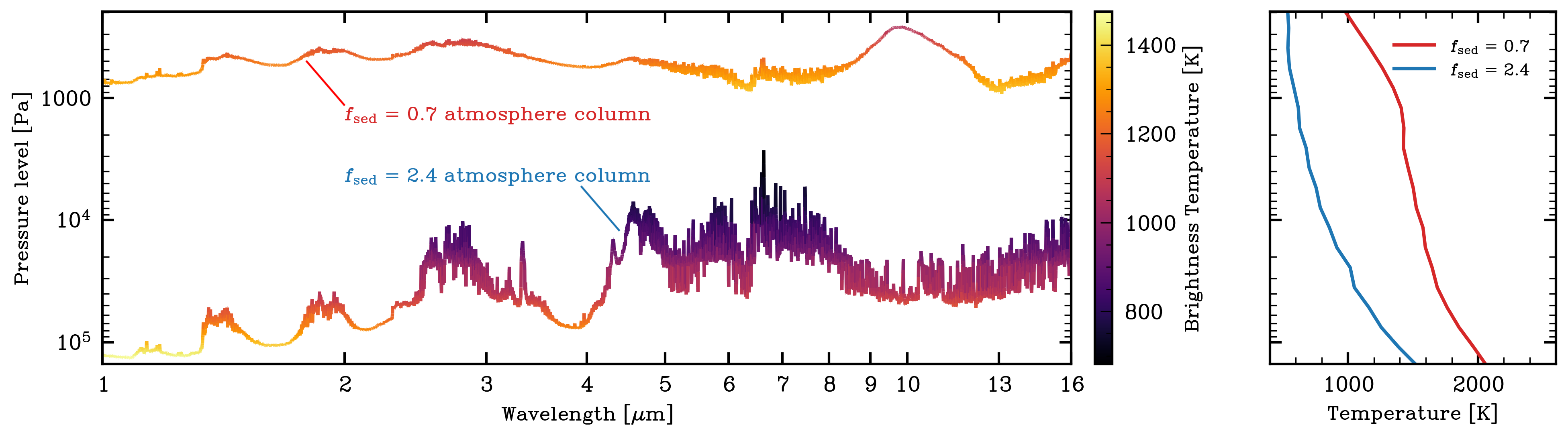}
    \caption{\textit{Left panel}: Photospheric pressures, where most of the thermal emission originates, for each component (atmosphere column) of the spectrum of \vhs; colour: temperature at the respective pressure. \textit{Right panel}: $P$--$T$ profile for the $\fsed=0.7$ (red) and $2.4$ (blue) column.}
    \label{fig: vhs contrib}
\end{figure*}

The molecular absorptions are well fitted owing to the $\fsed~=~2.4$ emission, while the silicate band is also adequately represented owing to the $\fsed = 0.7$ model which produces the necessary condition for silicate absorption at 10\,\mic. As seen in Fig.~\ref{fig: mmrs}, the low \fsed silicate cloud deck sits at higher altitudes (and therefore the $\taunuage = 1$ level) than the so-called photosphere, while the higher \fsed column has a thinner and lower-altitude cloud layer and thus a deeper $\taunuage = 1$ level. Fig.~\ref{fig: vhs contrib} shows that, as a result, the overall absorption in the $\fsed = 0.7$ column is dominated by clouds, leading to the outgoing flux emanating mostly from the cloud top at higher altitude, while the thin cloud coverage column radiates flux that is shaped by gas absorption in higher atmospheric layers than its thin cloud deck. The atmosphere in the $\fsed = 0.7$ column has a higher temperature than the former, owing to greenhouse effects, meaning that, although both \Teff values are the same, the overall temperature profile for $\fsed = 0.7$ is shifted by $\sim 500$\,K underneath the cloud deck of that column.

In both Fig.~\ref{fig: vhs best fit 1column} and Fig.~\ref{fig: vhs best fit} a small but distinct peak is apparent in the residuals at $\sim4.3$\,\mic, corresponding to an over-absorption in the model. This discrepancy is due to an excess of phosphine as a consequence of an incomplete understanding of the phosphorous chemistry (e.g., \citealp{Leggett2021}).

\section{Discussion}
\label{section: discussion}
\subsection[Implications for GJ 504 b and low-temperature objects]{Implications for GJ~504\,b and low-temperature objects}
The correction of the faulty D/H ratio for methane that had been applied in the previous \exo grid resulted in an appreciable revision of the derived parameters for GJ~504\,b. The \citet{Malin2025} study, that used the previous \exo grid and the same photometry, reported a higher effective temperature of $\Teff = 512^{+10}_{-10}$\,K and a lower surface gravity of $\logg = 3.45^{+0.35}_{-0.25}$ compared to our revised $\Teff = 473^{+14}_{-12}$\,K and $\logg = 4.0\pm 0.1$ (see Table\,\ref{tab:exorem_summary}). Critically, our higher surface gravity is more consistent with evolutionary models for its luminosity, resulting in a derived mass of $M = 6.0 ^{+1.7}_{-1.3}$\,\MJup and firmly placing GJ~504\,b in the younger planetary-mass regime. This finding, with the enriched C/O and [M/H], supports scenarios of planetary formation (e.g.\ core accretion) for this wide-orbit companion, contrasting with dynamical instability or gravitational fragmentation scenarios often invoked for BDs \citep{Pollack1996}. The magnitude of this parameter change highlights that the previous overabundance of $\text{CH}_3\text{D}$ in \exo biased the retrieved effective temperature and surface gravity for all \Teff$ \lesssim 700$K objects, where methane is highly abundant. We strongly recommend that all low-\Teff objects previously analyzed at methane-absorbing wavelengths using the older \exo grids be re-evaluated.

\subsection{The critical role of cloud thickness parametrisation}
The case of $\text{\vhs}$ brings into sharp perspective the need for a dimension geared at tuning cloud particle radii in substellar atmosphere grids, such as the \fsed parameter which was added to \exo. Grids relying on fixed cloud profiles or simple microphysics often fail to follow the overall shape of observed spectra, or access spectral features induced by highly optically thick clouds like the strong 10-\mic silicate absorption observed in \vhs and many other objects. We have demonstrated here that this feature can be accessed with \exoII models with a very low cloud sedimentation parameter $\fsed < 0.7$. A low \fsed ensures that the cloud layer remains suspended above the photosphere ($\tau \approx 1$ surface), allowing the cloud opacity ($\taunuage$) to dominate over the gas opacity, and thus giving rise to the necessary condition for silicate absorption. No other models have yet explored this low-\fsed regime: while the \texttt{SONORA-Diamondback} model includes \fsed parametrization, it does not extend to the extremely low values required to fully fit the silicate absorption in highly cloudy objects like $\text{\vhs}$, as evidenced by Fig.~\ref{fig: fixed fsed}. Furthermore, the \fsed parameter allows us to cover all objects on the CMD, no matter how red: simple microphysics follows the general trend in the L--T transition, but fails to represent the cloudiest of atmospheres. Our results suggest that \fsed could vary on BDs along the L--T transition, with the coolest L-dwarfs having low \fsed's, and warmest T-dwarfs having the higher \fsed's. A large-scale statistical analysis of \fsed values of BDs along the L--T transition derived through forward modelling could further confirm, and perhaps explain the mechanisms behind this behaviour.

\subsection[Towards the mapping of clouds on VHS 1256 b]{Towards the mapping of clouds on \vhs}
\label{subsection: towards the mapping of clouds on vhs}
Here we have shown that, with grids of pre-computed 1D models, one can nevertheless move towards a more complex description of objects with heterogeneous cloud cover. The relative success of the two-column model for \vhs further confirms that its atmosphere is better described as two distinct zones of differing cloud opacity, rather than one single homogeneous cloud coverage. Furthermore, with the lower \fsed exploration of \exoII and forward modelling framework, we managed to reproduce the silicate feature while fitting the overall spectrum and molecular absorptions (Fig.~\ref{fig: mol zooms}).

\begin{figure}[t]
    \centering
    \includegraphics[width=0.30\textwidth]{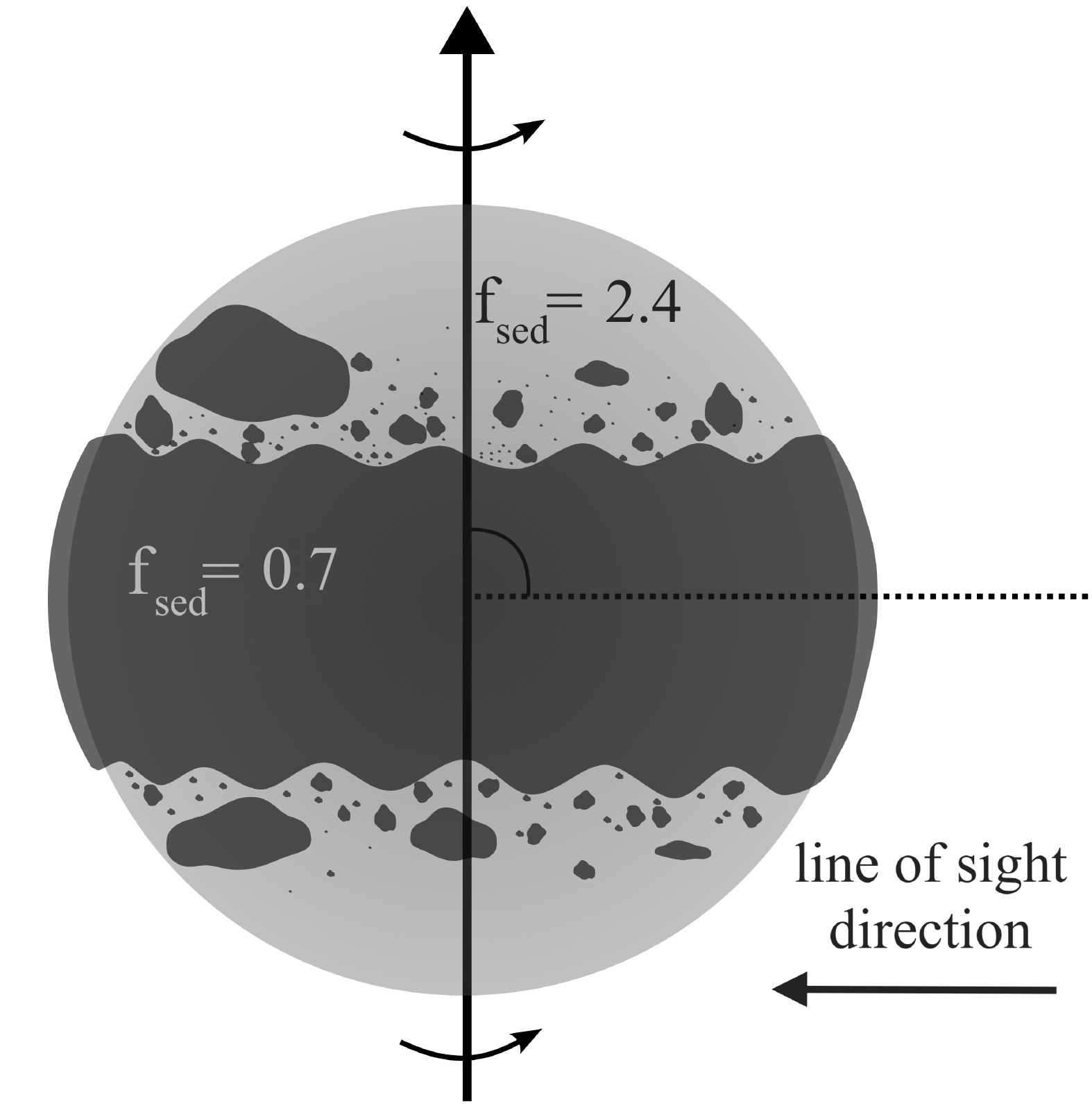}
    \caption{Simplified schematic of a possible cloud configuration of \vhs at the time the JWST observations were recorded. Here we show a physically plausible configuration of waves and spots that could cause the complex sinusoidal rotationally-modulated variations reported in \citet{Zhou2022}. The $\fsed = 0.7$ region would emanate flux from higher altitude and give rise to silicate absorption, while the $\fsed = 2.4$ region would emanate from lower regions, shaping predominantly the shorter wavelengths of the spectrum.}
    \label{fig: vhs diagram}
\end{figure}

With the two-column approach we can attempt to constrain thick-thin cloud surface fraction on patchy objects (here, the $\alpha$ parameter), provided that the viewing angle is constrained -- an equator-on view is favorable and more representative of the entire object's surface since variations are largely latitudinal \citep{Teinturier2026}. For \vhs, the $\alpha$ value we find implies a combination of 62\%\ thick cloud coverage of $\fsed = 0.7$, along with 38\%\ thinner cloud coverage of $\fsed = 2.4$, coherent with the large surface area of observable cloud expected for the equator-on view of this companion: the fraction we find seems qualitatively consistent with the GCM results in \citet{Teinturier2026} for a $\Teff = 1000$\,K brown dwarf, that predict a thick cloud belt at the equator. A possible configuration for this result is shown in Fig.~\ref{fig: vhs diagram}, with a cloud belt of $\fsed = 0.7$ as well as spots and zonal waves that could cause the elevated rotationally-modulated variability observed for \vhs. However, the $\alpha$ we find, in reality, only represents the proportions of the faces visible throughout the JWST observations. We expect the proportion of thick cloud coverage to change with time, causing the observed variability on \vhs.

Taking the simplistic solution where the object is strictly populated by the \fsed regimes ($\fsed = 0.7$ and 2.4), we can obtain a schematic view of how much the proportions of thick-to-thin cloud cover, $\alpha$, would need to change from one face of \vhs to the opposite, one face producing the minimum of the observed lightcurve and vice-versa (each with an associated $\alpha _{\mathrm{min}}$ and $\alpha _{\mathrm{max}}$ respectively). This gives us an order of magnitude of how much patchy cloud cover has structures that vary longitudinally. To do so, we compute the flux contrast between the two theoretical cloud regimes at the observationally time-resolved wavelengths and compare it to the actual observed amplitude in the lightcurve over a period of rotation:

For example, at the HST observation wavelength (1.27\,\mic), the two thick-thin cloud spectra proposed here for \vhs have a flux of $3.1\times 10^{-16}$ and $7.8\times 10^{-16}$\,W\,m$^{-2}\,\mic^{-1}$ respectively. This would produce a 96\% variation from the average disk-integrated flux at 1.27 \mic between faces over one rotation, if each hemisphere were completely populated by only thick or thin cloud coverage (so $\alpha = 0$ then $\alpha = 1$ half a rotation later; 100\% of the surface changing in \fsed value). From fitting the 2020 HST lightcurve  though, the object exhibits, over 1 rotation, an actual peak-to-peak amplitude of 24.7\% at 1.27\,\mic \citep{Bowler2020}, which requires that only a subset of the surface varies in \fsed value from one face to the other. Using this and the flux contrast between the two cloud components at 1.27\,\mic, we infer that a change of 25\% of the visible surface is sufficient to reproduce the HST observed modulation (so an $\alpha$ change of $\Delta \alpha = 0.25$). With the assumption that the $\alpha = 0.62$ we find in our two-column best fit lies at the average flux of the light curve at this wavelength, this corresponds to a thick cloud coverage changing from $\alpha_{\mathrm{thick}} = 0.50$ at the maximum to $\alpha_{\mathrm{thick}} = 0.75$ at the minimum flux on the HST lightcurve. 
\setlength{\tabcolsep}{2pt}
\renewcommand{\arraystretch}{1.45}
\begin{table*}
\caption{Summary of the retrieved atmospheric parameters for the \jwst \vhs data.}
\vspace{-0.3cm}
\label{tab:retrieval_results}
\centering
\small
\begin{tabular}[t]{lcccccccc}
\hline\hline
Method & Data used & 
\Teff [K] & 
\logg [dex]& 
$[{\rm M/H}]$ [dex]& 
C/O & 
\fsed (\fsedun)\tablefootmark{c} & 
\fseddeux & 
$\alpha$ \\
\hline
Classical\tablefootmark{a} & MIRI  
& $1038\pm3$ 
& $3.500\pm0.002$ 
& $1.41\pm0.04$ 
& $0.5500^{+0.0001}_{-0.0002}$ 
& $0.822\pm 0.003$ 
& $\cdots$ 
& $\cdots$ \\
Classical\tablefootmark{a} & \shortstack[t]{All}
& $1032.43\pm0.04$ 
& $4.1110\pm0.0004$ 
& $-0.1306\pm0.0002$ 
& $0.350001 ^{+0.000002}_{-0.000001}$
& $1.1958\pm0.0002$ 
& $\cdots$
& $\cdots$ \\

Two-column\tablefootmark{b} & \shortstack[t]{All}
& $1152.5\pm1.1$ 
& $4.00002^{+0.00027}_{-0.00022}$ 
& $0.0784^{+0.0021}_{-0.0019}$ 
& $0.6023^{+0.0006}_{-0.0007}$ 
& $0.70005^{+0.00013}_{-0.00008}$ 
& $2.443\pm 0.009$ 
& $0.6178^{+0.0028}_{-0.0030}$
\\
\hline
\end{tabular}
\tablefoot{%
The error bars listed here are only the statistical uncertainties from the nested sampling analysis; the misrepresentative nature of these is discussed in Section~\ref{subsection: posterior uncertainties}.
\tablefoottext{a}{Classical = one-column fitting (Section~\ref{subsubsection: classical approach})}
\tablefoottext{b}{
Section~\ref{subsubsection: two column approach}.}
\tablefoottext{c}{See Equation~(\ref{eq:Fcombo}) for \fsedun, \fseddeux, and $\alpha$.}
}
\end{table*}

With the same logic applied on the Spitzer observation of 5.76\% variation over a period at 4.5\mic \citep{Zhou2020}, we deduce a fraction of 6 \% longitudinal patchiness, so an alpha varying by $\Delta \alpha = 0.06$ between the face producing the maximum and that producing the minimum. The full calculations are detailed in Section~\ref{subsection: longitudinal frac calc}). The $\Delta \alpha$ values found at each wavelength are different, reflecting the overly-simplistic nature of this analysis -- while the two \fsed regimes we find produce a good fit to the spectrum, they can only go so far when describing the highly complex 3-D dynamics of \vhs (see the discussion in Section~\ref{subsection: advantages and limitations of the two column approach}). Furthermore, when attempting to constrain the cloud configuration of intensely variable objects, a huge caveat can lie in the nature of the observations themselves: in the test case of \vhs we have fitted the combined $\text{NIRSpec}$ and $\text{MIRI}$ spectrum as a single, instantaneous snapshot. However, \vhs has a rotational period of $\sim 22$ hours, and is dramatically time-variable (the order of magnitude is shown as red error bars on Fig.~\ref{fig: vhs best fit}). Given the time elapsed between the $\text{NIRSpec}$ and $\text{MIRI}$ observations, it completed a significant fraction of its rotation, meaning that the observed surface, and thus the overall cloud fraction $\alpha$, likely varied temporarily during the measurement -- we are fitting a time-averaged spectrum with a single $\alpha$ fraction, that would in fact be changing as the planet rotates. This introduces additional sources of uncertainties that are not taken into account in the fit, and highlights that our two-column solution, while highly successful spatially, cannot account for the full spatio-temporal dynamics of the atmosphere. 

Here, with a simple two-column approach, we have derived the rough proportions, thickness, and even variation of cloud coverage with time, but we cannot constrain the shape, nature and exact distribution of the structures on \vhs: the HST lightcurve has complex variations and \citet{Zhou2022}, thanks to lightcurve fitting, show that it is best represented by three sinusoidal components with periods of $\sim 19$, 15 and 10.5\,h, suggesting multiple spots and waves modulating the variability, each with much smaller peak-to-peak amplitudes than that found by \citet{Bowler2020}. \citet{Tan2025}, with their GCM, found that the heterogeneities on \vhs could come primarily in the form of a complex massive planetary-scale silicate and iron dust storm creating cloud radiative feedback, rotating eastwards in and out of view. However, their lack of disequilibrium chemistry is a limitation, and, with our inclusion of disequillibrium chemistry and finer grid, which is more computationally achievable for 1D models, we obtain a substantially closer fit of the spectral features on the JWST spectrum, in particular for the CO, H$_2$O, and CH$_4$ bands, as Fig.~\ref{fig: mol zooms} shows.

Multi-wavelength time-resolved monitoring is the most effective pathway to constrain the full $\text{3D}$ heterogeneous structure of sub-stellar objects, and obtaining more for \vhs will be decisive in corroborating or correcting any of the preliminary theories put forward here regarding its atmospheric configuration. To further constrain the vertical structure of clouds on \vhs, we necessitate time-resolved monitoring at adjacent wavelengths: since different wavelengths probe distinct atmospheric heights, if the modulation is weaker at the wavelength probing the higher atmosphere compared to the modulation of the latter, this could indicate an intermediate cloud deck between both layers. The modulations can be directly compared if the two wavelengths probed are closeby and thus have almost identical cloud optical properties and Planck function \citep{Zhou2020}, such as the 1.4\,\mic continuum region with its adjacent water absorption that probes much higher altitudes. From our analysis, we would expect an anti-correlation between simultaneous observations at 1.27~and 4.5\,\mic, with the flux being higher in the thinner cloud zone at 1.27\,\mic and the opposite at 4.5\,\mic; simultaneous time-resolved monitoring at these wavelengths will be decisive in further refining our knowledge of \vhs.

\subsection{Advantages and limitations of the two-column approach}
\label{subsection: advantages and limitations of the two column approach}
The two-column approach allows for a non-homogeneous description of atmospheres in cases where patchy cloud cover is strongly suspected, and where, as a result, a single 1D model does not fit. This constitutes an excellent intermediate approach from GCMs which are time-consuming and, as it stands, cannot be generated for each object or made into large grids geared at closely fitting any observation. However, this approach entails making extreme simplifications.

An assumption made is that the atmosphere is described only by two distinct vertical cloud distributions, with a sharp transition from thicker to thinner cloud zones rather than the predicted (albeit steep) gradient from thinner to thicker cloud cover shown in GCMs \citep{Teinturier2026}. Since our methodology adopts individual thermal profiles for each of the two atmospheric columns, it implicitly assumes no horizontal heat redistribution between them; at the other extreme, other studies (e.g.\ \citealt{Marley2010, Morley2014a, Zhang2025, Mollire2025}) converge to a single thermal profile for the entire atmosphere, thereby implying complete heat redistribution. In reality, an intermediate regime is expected, in which the degree of heat redistribution is governed by the competition between radiative cooling timescales and dynamical transport timescales. In the deep atmosphere, at pressures exceeding $\sim10^{6}$\,Pa, the horizontal advective timescale is expected to be shorter than the radiative timescale, resulting in efficient thermal homogenisation and similar temperature structures between cloudy and non-cloudy regions. At lower pressures, however, the radiative timescale decreases rapidly and becomes shorter than horizontal advective timescales, allowing local radiative equilibrium to dominate and thermal contrasts to persist between regions of differing cloud opacity \citep{Seager2010}. The atmospheric layers probed by the \vhs spectrum ($\sim10^3$--$10^{5}$\,Pa) lie firmly within this latter regime; using $\tauadv = R_{\rm p}/v$ with $R_{\rm p}$ the planet radius and $v$ the horizontal wind speed ($\sim~100$\,m\,s$^{-1}$; \citealp{Teinturier2026}), and $\taurad=Pc_{\rm p}/4g\sigma T^3$ with $P$, $T$, $c_{\rm p}$, $g$ and $\sigma$ the pressure, temperature, specific heat capacity, gravity, and Stefan--Boltzmann constant, we find $\taurad \sim 3 \times 10^4$\,s and $\tauadv \sim 7 \times 10^5$\,s, so $\taurad \ll \tauadv$. Therefore, at these pressures, horizontal heat redistribution is weak and a distinct temperature difference (of $\sim 300$\,K) between cloudy and cloud-free zones is predicted by the GCM presented by \citet{Teinturier2026}. 

However, our thick- and thin-cloud thermal profiles in Fig.~\ref{fig: vhs contrib} exhibit a more pronounced divergence ($\sim500$\,K) within the layers probed by the observations. It is improbable that this temperature contrast could physically be maintained between the centres of thick and thin cloud zones at these pressures; this is a major limitation in our method. The vast thermal contrast we find would likely be reduced if the effective temperatures between both zones were also allowed to vary, a physical attribute demonstrated by GCM studies, which showed an expected difference in outgoing longwave radiation between cloudy and non-cloudy zones owing to the reduction of emergent flux from thicker cloud layers. However this entails adding a dimension to the forward modelling analysis that is already time-consuming due to the high-resolution and expanse of the JWST data used here. An addition to our framework whereby temperature contrasts beyond a certain point are forbidden could address the problem of large thermal contrast, but would require a physically motivated threshold of temperature difference; although \citet{Teinturier2026} predict a value, discrepancy also arises from the fact that their GCM is computed under different conditions, without the inclusion of iron clouds which are expected to be present at these temperatures and would enhance the greenhouse effect, increase local radiative timescales, and thus exacerbate temperature contrasts between cloudy and non-cloudy regions; they also adopt $\log g = 5$\,dex and a 5\,h rotation period. \vhs, with its slow rotation coupled with unusually high variability, is expected to display substantially different atmospheric dynamics. 

The \exo model has the functionality of simulating patchy clouds, similarly to \citet{Marley2010}, as a linear combination of a cloudless and cloudy column and evaluating the radiative transfer with one single thermal profile. However, doing so would lead to an additional dimension (cloud fraction) to the computed grid of models, and would require choosing one \fsed value for each column, thus over-simplifying the cloudy scheme. Instead, the approach chosen here allows for any combination of \fsed instead of assuming completely cloudy and cloudless patches. On the other hand, two-column retrieval frameworks require on-the-fly calculations of both atmospheric chemistry and radiative transfer, enabling a highly flexible, data-driven exploration of cloud properties. Such an approach allows for fine-tuning across a wide range of cloud parameters, including $\fsed$, $\Kzz$, particle size distributions, and cloud mass fractions at the base of the cloud deck. Applying a similar retrieval framework to \vhs, in which multiple silicate cloud species are explored, could plausibly yield an even closer match to the observed 10\,\mic feature \citep{Hoch2025}. The forward-modelling approach adopted here, however, relies exclusively on pre-computed grids of atmospheric models and is therefore considerably less computationally demanding; in fact many retrieval analyses, due to their high numeral cost, are constrained to downgrading the observation resolution (e.g.\ by \citealt{Zhang2025}), thus losing precious spectral information.

Rather than aiming to recover a fully self-consistent and physically accurate description of the atmosphere, our objective is to demonstrate that combining multiple 1D atmospheric columns with differing cloud properties provides a practical and physically motivated step towards capturing the spectral complexity of patchy atmospheres. This approach enables a more nuanced representation of cloud heterogeneity than a single 1D model, while avoiding the need to run sophisticated GCMs, perform on-the-fly radiative transfer, or introduce additional dimensions into already memory-intensive model grids.

\section{Conclusions}
\label{section: conclusions}
Here we have taken measures to step towards a more nuanced and complex description of the atmospheres of substellar objects when using pre-computed 1-D models, in particular focusing on the modelling of their cloud coverage. To this end, we have: 
\begin{enumerate}
    \item generated a new, coherent set of self-consistent \exo atmosphere models (at $R=500$ and $R=\textrm{10,000}$), titled \exoII, with updated molecular opacities, most notably changing the alkali line profiles for experimentally-validated broader line wings. Critically, the implementation of robust numerical convergence criteria ensures the final grids are free from unphysical, numerically unstable spectra that were plaguing the previous \exo grid \citep{Charnay2018};
    
    \item corrected the $\text{CH}_3\text{D}$ overabundance in the previous grid, which has resulted in an appreciable revision of the derived parameters for GJ~504\,b, yielding a lower effective temperature ($\Delta \Teff = 39$~K; $\sim -3\sigma$) of $\Teff = 473^{+14}_{-12}$\,K and a higher surface gravity ($\Delta \logg = 0.55$; $\sim +7 \sigma$) of $\logg = 4.0\pm 0.1$\,dex than in the previous study using \exo with the same data \citep{Malin2025}. This finding places the companion firmly in the young, planetary-mass regime ($M = 5.4 ^{+1.5}_{-1.4}$\,\MJup). A re-evaluation of all low-temperature objects previously modelled with the older $\text{Exo-REM}$ grid is advised;
    
    \item included the sedimentation parameter, ranging from  $0.5~\leq~\fsed~\leq 9$, which proves essential for modelling the extreme range of cloud opacities in substellar atmospheres. We demonstrated that low \fsed values ($<1$) are necessary to access the strong silicate absorption characteristic of highly optically thick clouds, a regime not fully explored by previous models. We confirm that \fsed must evolve, increasing across the L--T transition to explain the sinking cloud deck;
    
    \item introduced and applied a two-column forward modelling scheme for pre-computed grids of self-consistent 1D models, which emulates cloud patchiness by linearly combining two self-consistent atmosphere columns differing in cloud optical thickness and thermal profiles. This simplistic approach constitutes a vast improvement from using singular 1-D models on patchy objects, whilst bipassing the need to add dimensions to pre-computed grids or to run computationally costly retrievals or GCMs;
    
    \item put forward a patchy cloud configuration for $\text{\vhs}$ with the two-column approach, considerably improving the fit of the full $\text{JWST}$ spectrum compared to the classical 1D approach. The best fit model sums a $62\%$ thick cloud component ($\fsed = 0.7$) and a $38\%$ thin cloud component ($\fsed = 2.4$), with the strong silicate absorption being produced by the high-altitude and opacity of the low-\fsed column. The derived bulk parameters are $\Teff = 1153$\,K, $\logg = 4.0$\,dex, [M/H] = 0.08\,dex, and $\text{C/O} = 0.60$.
\end{enumerate}
This work establishes the next generation of $\texttt{Exo-REM}$ grids as a robust and accessible framework for interpreting directly imaged brown dwarf and giant planet photometry and spectroscopy. A high-resolution \exoII grid\footnote{\url{https://lesia.obspm.fr/exorem/YGP_grids/Exo-REMk26/High_Res_grid_2026/R200k_cloudyfsed_2026/}} ($R = \textrm{200,000}$) spanning 0.9--6.0\,\mic necessary for detailed chemical and kinematic studies from high-resolution spectrographs (e.g.\ VLT/CRIRES$^+$, VLT/HiRISE) is already available and will be presented in an upcoming publication (Radcliffe et al., in prep.).

\begin{acknowledgements}
This project has received funding from the European Research Council (ERC) under the European Union’s Horizon 2020 research and innovation programme (COBREX; grant agreement No.~885593). This work was granted access to the HPC resources of MesoPSL, financed by the \^{I}le-de-France Region and the Equip@Meso project (reference ANR-10-EQPX-29-01) of the \textit{Investissements d'Avenir} programme supervised by the Agence Nationale pour la Recherche.
S.P.\ was supported by an appointment to the NASA Postdoctoral Program at the NASA-Goddard Space Flight Center, administered by Oak Ridge Associated Universities under contract with NASA.
G.-D.M.\ acknowledges the partial support of the Deutsche Forschungsgemeinschaft (DFG) through grant MA~9185/2-1.
\end{acknowledgements}


\bibliographystyle{aa_url}
\bibliography{PhD}


\begin{appendix}

\section[The Exo-REM k26 model]{The \exoII model}
This section provides additional information on the \exoII model.

\subsection{Example volume mixing ratio profiles}
\begin{figure}[H]
    \centering
    \includegraphics[width=\largimgunecol]{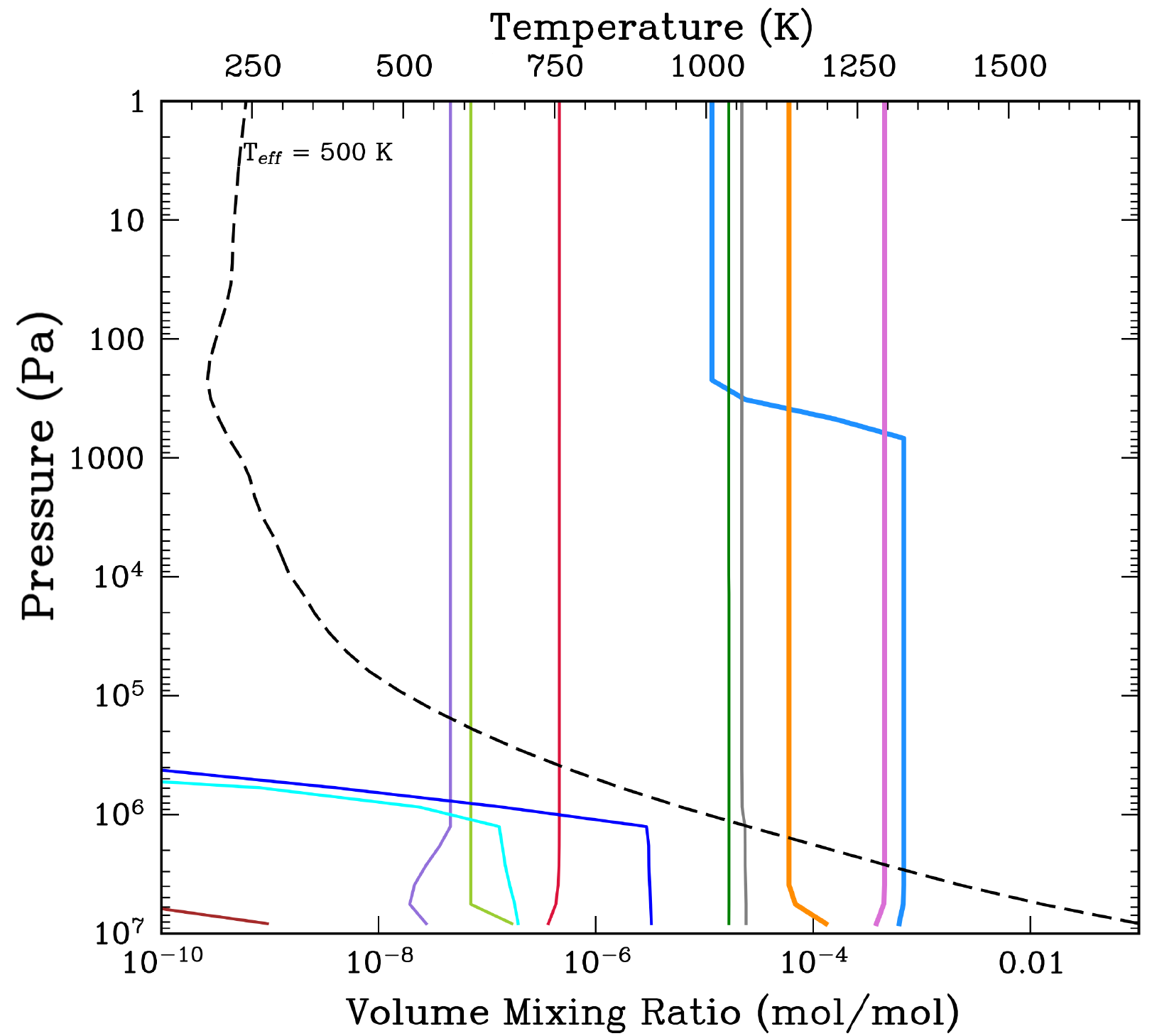}
    \includegraphics[width=\largimgunecol]{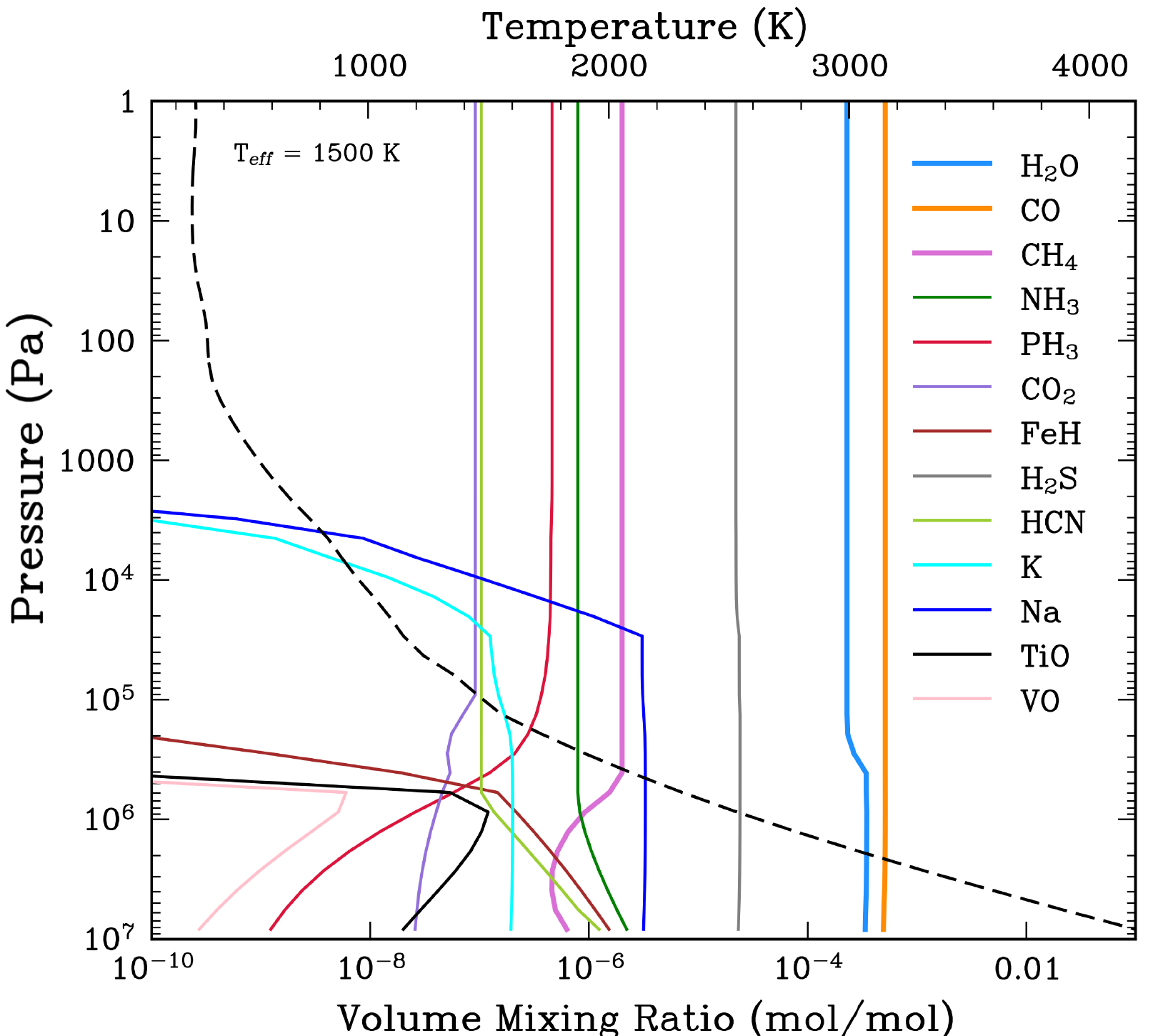}
\caption{Volume mixing ratio (VMR) profiles for all the molecules included in the atmosphere, including non-equilibrium chemistry. These are computed for $\Teff = 500$~K (top panel) and 1500~K (bottom panel), $\logg = 4$ and solar C/O and [M/H] values. H$_2$ and He are not plotted here as they constitute the background and thus the majority of the atmosphere.}
\label{fig:vmrs}
\end{figure}

Fig.~\ref{fig:vmrs} provides volume mixing ratio profiles (the abundances of each species) at different effective temperatures, for the cloudless model.

\subsection{Line lists and abundances}

\begin{figure}[!t]
    \centering
    \includegraphics[width=\largimgunecol]{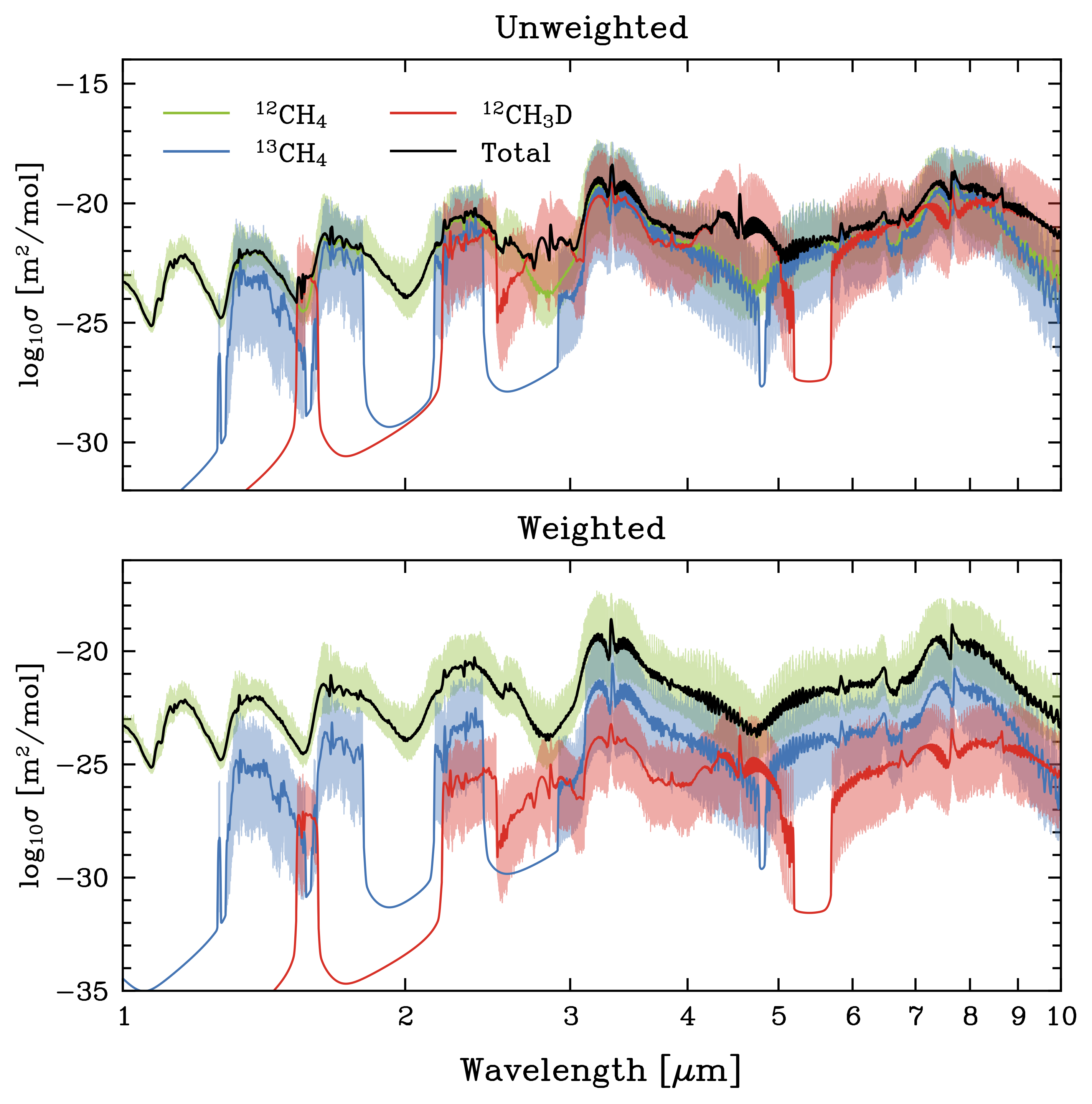}
    \caption{Methane isotopologue absorption cross sections at $P = 1$\,bar, $T = 500$\,K, discussed in Section~\ref{subsubsection: line lists}. The translucent lines correspond to the $R = 10^6$ cross sections, while the thick and opaque ones are those same cross sections binned down to $R = 500$. The top panel shows the cross sections for equal abundances of each isotopologue (as was erroneously implemented previously for CH$_3$D in \exo), while the bottom panel shows the cross sections when weighted according to solar abundances (i.e. D/H ratio of $2\times10^{-5}$).}
    \label{fig: ch4_iso}
\end{figure}

We provide here the detailed list of abundances and line lists in Table~\ref{tab:isotopologues} referred to in Section~\ref{subsubsection: line lists}, as well as the element abundances in Table~\ref{tab:isotopes} from \cite{Asplund2021} and an illustration of the error in the previous \exo grid with the proportion of deuterium in the methane isotopologue mixture in Fig.~\ref{fig: ch4_iso}.

\renewcommand{\arraystretch}{1.2}
\begin{table}[H]
\centering
\caption{Abundances and line lists for selected species.}
\footnotesize
\setlength{\tabcolsep}{4pt}
\begin{tabular}{llll}
\hline \hline
Molecule & Isotopologue & Fraction & Line List \\
\hline
CH$_4$  & $^{12}$CH$_4$  & 9.8885e-1 & HiTEMP (1) \\
        & $^{12}$CH$_3$D & 7.9110e-5 & HiTRAN (2) \\
        & $^{13}$CH$_4$  & 1.1069e-2 & HiTRAN (3) \\
\noalign{\smallskip}
CO      & $^{12}$C$^{16}$O & 9.8616e-1 & HiTEMP (4)\\
        & $^{12}$C$^{17}$O & 3.5960e-4 & HiTRAN (3) \\
        & $^{12}$C$^{18}$O & 1.8660e-3 & HiTRAN (3)\\
        & $^{13}$C$^{16}$O & 1.0944e-2 & HiTRAN (3) \\
        & $^{13}$C$^{17}$O & 3.9800e-6 & HiTRAN (3) \\
        & $^{13}$C$^{18}$O & 2.0870e-5 & HiTRAN (3)\\
\noalign{\smallskip}
H$_2$O  & H$_2^{16}$O & 9.9752e-1 & HiTRAN (3) \\
        & H$_2^{17}$O & 3.5960e-4 & HiTRAN (3) \\
        & H$_2^{18}$O & 1.8660e-3 & HiTRAN (3) \\
        & HD$^{16}$O  & 4.0008e-5 & HiTEMP (3) \\
        & HD$^{17}$O  & 1.4384e-7 & HiTEMP (3) \\
        & HD$^{18}$O  & 7.4624e-7 & HiTEMP (3)\\
\noalign{\smallskip}
NH$_3$  & $^{14}$NH$_3$ & 9.9750e-1 & ExoMol (8) \\
        & $^{15}$NH$_3$ & 2.1750e-3 & HiTRAN (2) \\
\noalign{\smallskip}
TiO     & $^{46}$TiO & 8.2245e-2 & ExoMol (10) \\
        & $^{47}$TiO & 7.4323e-2 & ExoMol (10) \\
        & $^{48}$TiO & 7.3586e-1 & ExoMol (10) \\
        & $^{49}$TiO & 5.3526e-2 & ExoMol (10) \\
        & $^{50}$TiO & 5.0908e-2 & ExoMol (10) \\
\noalign{\smallskip}
CO$_2$  & $^{12}$C$^{16}$O$_2$ & 1 & HiTEMP (4) \\
FeH     & $^{56}$Fe$^{1}$H & 1 & ExoMol (5) \\
H$_2$S  & $^{1}$H$_2^{32}$S & 1 & HiTRAN (2) \\
HCN     & $^{1}$H$^{12}$C$^{14}$N & 1 & ExoMol (6) \\
$^{39}$K & $^{39}$K & 1 & Allard (7) \\
$^{23}$Na & $^{23}$Na & 1 & Allard (7) \\
PH$_3$  & $^{31}$P$^{1}$H$_3$ & 1 & ExoMol (9) \\
VO      & $^{51}$V$^{16}$O & 1 & ExoMol (10) \\
\hline
\end{tabular}
\label{tab:isotopologues}
\vspace{+0.15cm}
\begin{minipage}{\linewidth}
{\footnotesize \textbf{References}: (1)~\citet{Hargreaves2020}. (2)~\citet{Rothman2013}. (3)~\citet{Gordon2022}. (4)~\citet{Rothman2010}. (5)~\citet{Bernath2020}. (6)~\citet{Harris2006}. (7)~\citet{Allard2016, Allard2019}. (8)~\citet{Coles2019}. (9)~\citet{sousa-silva2015}. (10)~\citet{McKemmish2019}. (11)~\citet{McKemmish2016}.}
\end{minipage}
\vspace{-0.2cm}
\end{table}

\begin{table}[t]
\centering
\caption{Isotopic abundances (by number) of selected elements. Quantities derived from \citet{Asplund2021}.}
\begin{tabular}{cccc}
\hline \hline
$Z$ & Element & $A$ & Fraction \\
\hline
1  & H   & 1  & 0.99998  \\
1  & H   & 2  & 0.00002 \\
6  & C   & 12 & 0.98893 \\
6  & C   & 13 & 0.01107 \\
7  & N   & 14 & 0.99775 \\
7  & N   & 15 & 0.00225 \\
8  & O   & 16 & 0.99776 \\
8  & O   & 17 & 0.00036 \\
8  & O   & 18 & 0.00188 \\
22 & Ti  & 46 & 0.08250 \\
22 & Ti  & 47 & 0.07440 \\
22 & Ti  & 48 & 0.73720 \\
22 & Ti  & 49 & 0.05410 \\
22 & Ti  & 50 & 0.05180 \\
\hline
\end{tabular}
\label{tab:isotopes}
\end{table}

\subsection{Model convergence}
\label{subsection: robust convergence criteria}
Here we present in more detail the process of flagging and eliminating unconverged spectra in the \exoII model.

\subsubsection{Flags for unconverged gridpoints}

The following are the different types of cases that we considered as unconverged gridpoints:
\begin{enumerate}
 \item Cases which were so unstable that the algorithm crashed, and did not produce an output spectrum or $P$--$T$ profile. These most likely occur when the input $P$--$T$ profile is too different from the final solution, and are easily flagged since they manifest simply as a missing gridpoint.

 \item Cases where the output spectrum did not obey the Stefan--Boltzmann law; in a one-dimensional model assuming radiative-convective equilibrium, the net flux $F$ -- comprising both radiative and convective components -- should be constant across different pressure levels and equal to
\begin{equation}
\label{eq: total flux}
    F = \sigma \Teff^4.
\end{equation}
We integrated the spectrum over the wavelength range of the model and considered as unconverged those cases for which the total flux deviated by more than 1\%\ from the expected total flux in Equation~\ref{eq: total flux}. 

 \item Finally, some cases may pass criteria 1.\ and 2.\ but have an associated $P$--$T$ profile that contains a large temperature inversion, a drop in temperature with increasing pressure. The thermal structure of a BD/YGP is convective in the deeper layers of its atmosphere, where high electron densities prevent thermal photons from travelling long distances, making convection the primary mode of energy transport \citep{Marley2015}. The gradient of a $P$--$T$ profile in these dense atmosphere zones is taken to follow closely

 the convective adiabat \citep{Baraffe2002}, and would not contain any temperature inversions. Eventually, as radiation removes more energy from the thinning atmosphere, the temperature gradient with altitude becomes less steep, marking the end of the convective region; here, the thermal profile is governed by radiative equilibrium \citep{Marley2015}, tending towards an isotherm at the top layers. There may be small temperature inversions at these layers, but one would not expect to see any dramatic inversions, especially in the deeper, convective layers. Thus, a large temperature inversion beyond the first few layers is a robust marker for an unconverged \exo spectrum, as  it is unphysical and only numerically arises from a difficulty for the code to iteratively reach the correct input \Teff and total flux. Very high temperatures are required for thermal inversions ($>2500$\,K), and these occur in the radiative zone \citep{Madhusudhan2019}.
\end{enumerate}

\subsubsection{Treatment of unconverged gridpoints}
To minimise the number of missing gridpoints after a first run of \exo, we have developed a new procedure for the flagged simulations. For those that have crashed (1.), we try again with an adjacent $P$--$T$ profile which, in practice, is one with the same input parameters and a C/O ratio $\pm$ 0.05. We choose to slightly alter this parameter as it is the one that has the least effect on resulting $P$--$T$ profiles and spectra. Furthermore, we subdivide category~2.\ into spectra deviating between 1-2\% from the total expected flux, and those deviating more than 2\%; for the former, we deem that the profiles must not be far from convergence and we rerun \exo with the output $P$--$T$ profile as a new input, giving it more iterations to converge. For the latter case, we select, just like for category~1., an adjacent $P$--$T$ profile with $\co \pm 0.05$ and rerun. Lastly, for category~3.\ where an inversion is detected in the final $T$--$P$ profile, we also inject an adjacent $P$--$T$ profile and start from scratch. In all of these cases, the majority of gridpoints were able to reach convergence, and those that did not are discarded to avoid biasing the grid with unphysical points. These constituted 0.25\%\ of the cloudless grid, 0.1\%\ of the cloudy microphysics grid and 3.4\%\ of the \fsed grid and are listed on the \exo site. The most problematic zones were the high \Teff and low \fsed values as they were the most unstable.

The most cloudy cases had the largest number of gridpoints that were not able to converge: this is due to the unstable nature of adding cloud physics to the model. For every iteration of the algorithm, the chemistry is computed with an associated $P$--$T$ profile; only then are the clouds added, which affects in turn, the thermal structure. However, the chemistry is not recomputed once the clouds are added, and thus if they had a strong impact on the thermal structure, the next iteration will have a drastically different associated profile. As a result, the final spectrum may conserve flux, but it is possible that not each level of the atmosphere is converged, leading to a $T$--$P$ inversion (category~3.). 

\subsection[Exo-REM k26 CMD with varying fsed]{\exoII CMD with varying \fsed}
\begin{figure}[t]
    \centering
    \includegraphics[width=\largimgunecol]{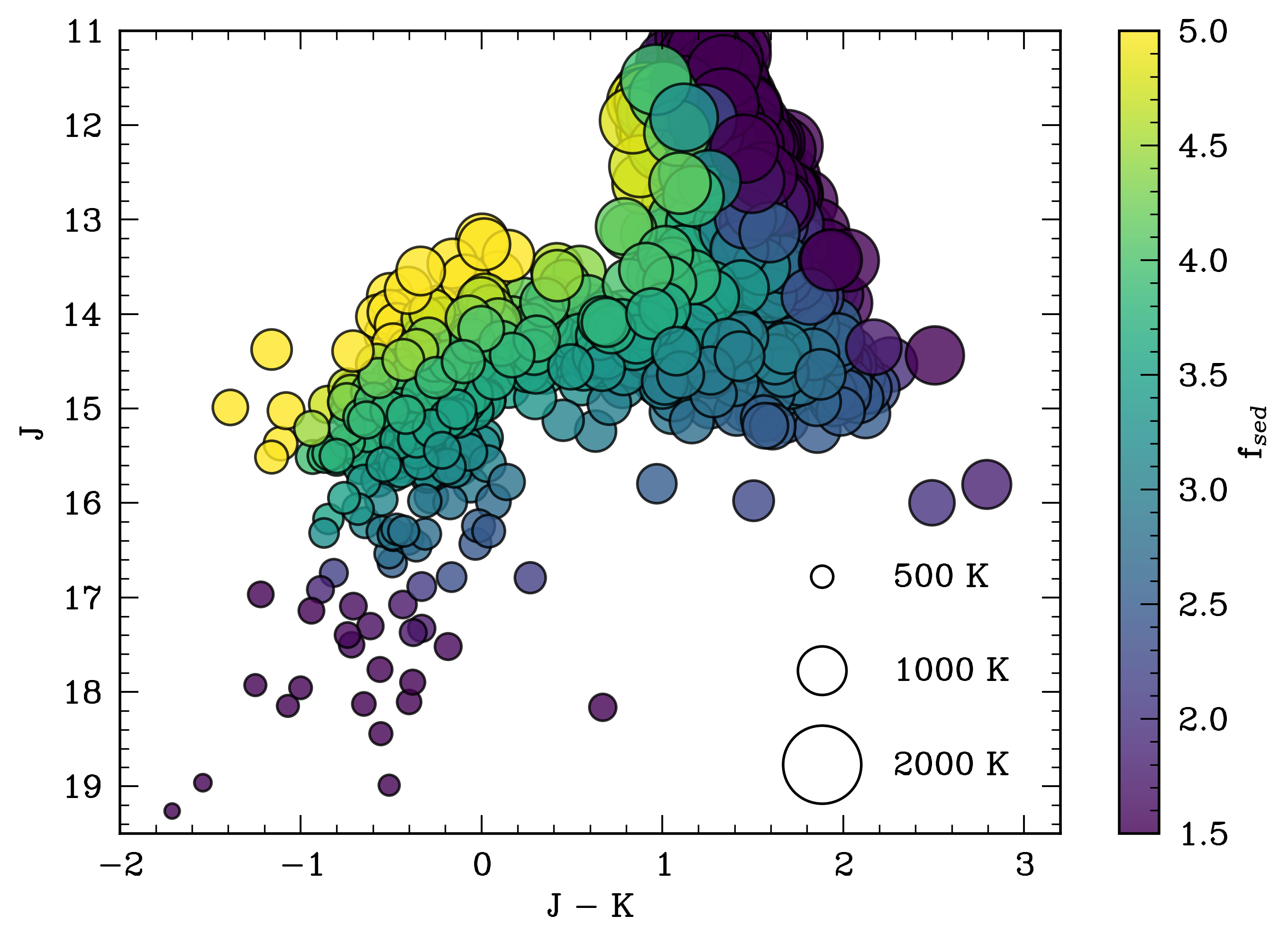}
    \caption{CMD with the data points sized according to their interpolated \Teff, and coloured by their interpolated \fsed found by computing the magnitudes of \exoII spectra of varying \fsed and \Teff values at $\logg = 4.5$ and solar C/O and [M/H], then interpolating to find the \fsed and \Teff at the \citet{Best2025} data points. This highlights the \fsed trend discussed in Section~\ref{subsection: reproducing the l--t transition}}
    \label{fig:lt_fsed2}
\end{figure}

Fig.~\ref{fig:lt_fsed2}
shows the \fsed trend along the L--T transition discussed in Section~\ref{subsection: reproducing the l--t transition}.

\section{Posterior distributions}
In \ref{subsection: posterior uncertainties} we discuss the error bars in the posterior distributions from the forward modelling carried out on \vhs in Section~\ref{subsection: vhs 1256b} and supply all of the posterior distributions for GJ~504~b and \vhs:
\subsection[Reduced chi-square and posterior uncertainties in the \vhs analysis]{\chisqreduced and posterior uncertainties in the \vhs analysis}
\label{subsection: posterior uncertainties}
The Bayesian framework propagates the input data error in the posterior distributions: since the input JWST data for \vhs has minute error bars (the average across the NIRSpec and MRS data being $\sim$5\%), the posteriors are narrow and with a dispersion that does not account for model systematics. The \chisqreduced of the best fitting models found here, while being considerably smaller than those found in previous studies on the same data, have values much larger than 1, which indicates that errors other than noise dominate the error budget, such as additional flux errors in the JWST data and, more importantly, modelling errors. We carried out an identical run where we multiplied the NIRSpec and MIRI noise error by a factor of 10.6 to bring the value of \chisqreduced for the best fitting model to 1.0. This is similar to the approach of \citet{Gibson2020}, who introduced a $\beta$ parameter to scale the noise amplitude and modified accordingly the likelihood function, as well as \citet{Zhang2025} who added noise, through an hyper-parameter $\log b$, to their NIRSpec data. Our test, as intended, brought the final \chisqreduced to 1 and inflated the $\pm 1\sigma$ values of each posterior distribution: each parameter went from an error of $\sim$0.02\% to $\sim$0.2\%, with the exception \Teff that increased from $\sim$0.06\% to $\sim$0.9\%. Although this had the effect of amplifying the $\pm 1\sigma$ values by a factor of $\sim$10, these are nonetheless still misrepresentative of the actual error on this analysis that is dominated by model systematics, and we show the original, non-inflated distributions in Figs.~\ref{fig: vhs best fit MIRI}, \ref{fig: vhs best fit 1column} and~\ref{fig: vhs best fit}. Furthermore, the solutions tend to converge to grid nodes, (e.g.\ [M/H] = 1.50\,dex, and C/O = 0.55 for the MIRI/MRS fit). This is a recurring issue when using pre-computed grids, whereby gridpoints that are synthesised directly by the model will sometimes have appreciably higher likelihoods than the adjacent interpolated spectra.

\subsection{Corner plots}
\begin{figure*}[!t]
    \centering
    \includegraphics[width=\largimgdeuxcols]{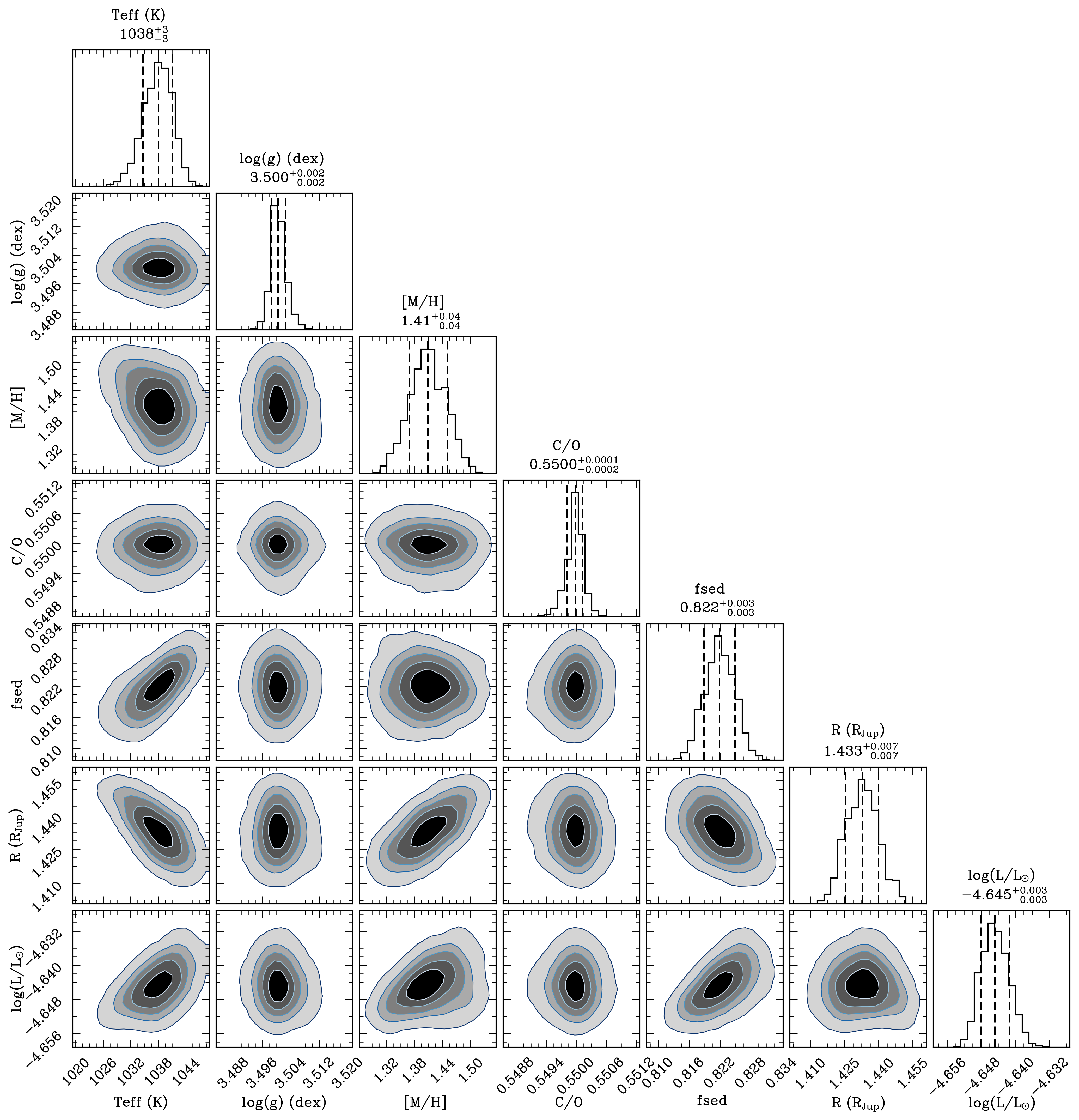}
    \caption{Posterior distribution from \vhs forward modelling with \texttt{ForMoSA} using the medium-resolution \exoII grid and just MIRI/MRS observations, with only one column (so the classical approach). The 3$\sigma$, $2\sigma$, 1.5$\sigma$, $1\sigma$, and 0.5$\sigma$ regions are shown in regions shaded from grey to black (in respective order), and the listed uncertainties correspond to 1$\sigma$. The corresponding best fit is shown in Fig.~\ref{fig: vhs best fit MIRI}. Flat priors are given spanning the whole \exoII grid range detailed in Table~\ref{tab: models}.}
    \label{fig: vhs mrs corner}
\end{figure*}

\begin{figure*}[!t]
    \centering
    \includegraphics[width=\largimgdeuxcols]{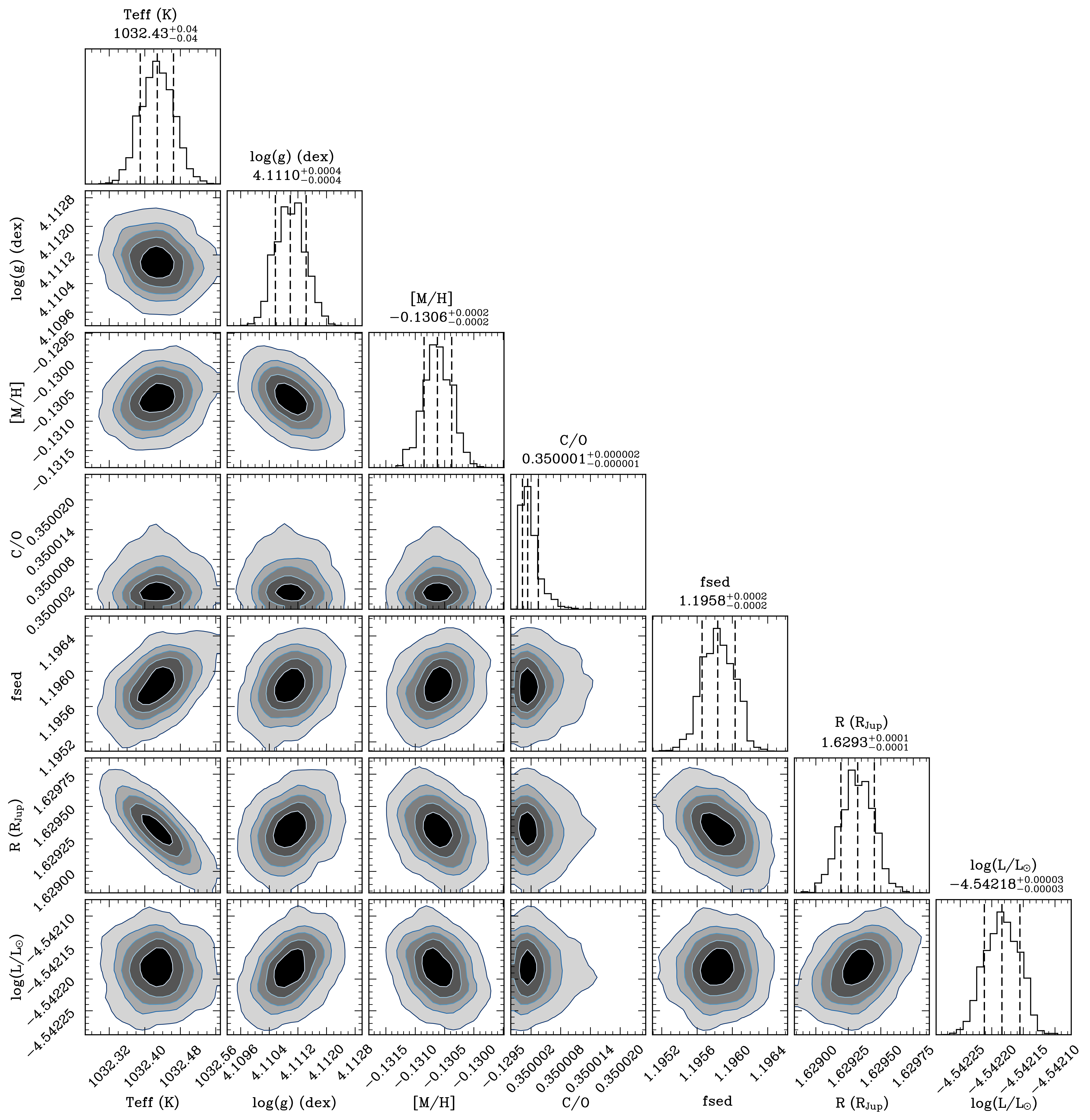}
    \caption{Same as Fig.~\ref{fig: vhs mrs corner} but using all of the JWST observations. The corresponding best fit is shown in Fig.~\ref{fig: vhs best fit 1column}.}
    \label{fig: vhs all obs corner}
\end{figure*}

\begin{figure*}[!t]
    \centering
    \includegraphics[width=\largimgdeuxcols]{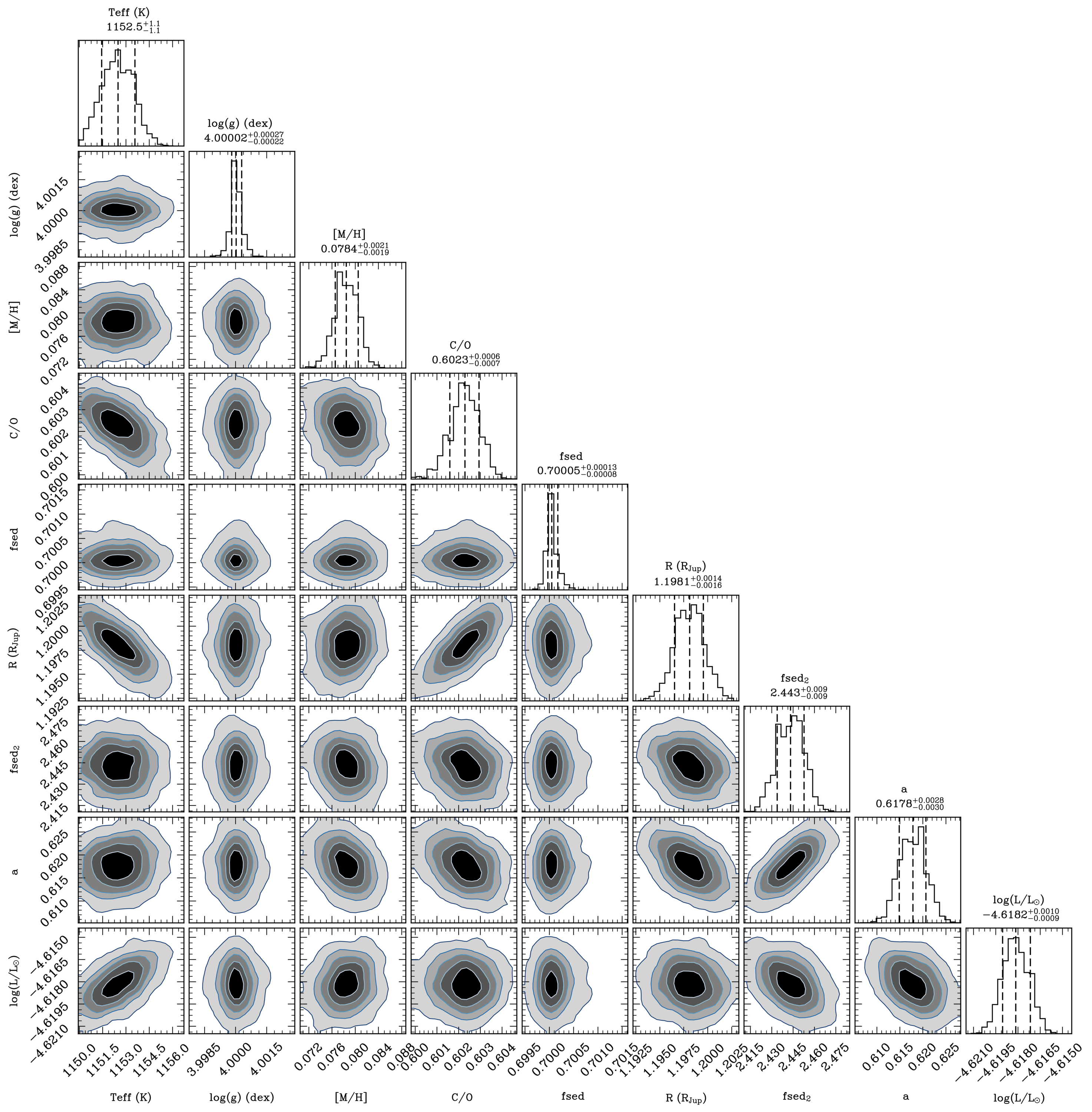}
    \caption{Same as Fig.~\ref{fig: vhs all obs corner}, but in a two-column framework as detailed in Section~\ref{subsubsection: two column approach}. The corresponding best fit is shown in Fig.~\ref{fig: vhs best fit}.}
    \label{fig: 2fsed allobs vhs corner}
\end{figure*}

\begin{figure*}[!t]
    \centering
    \includegraphics[width=\largimgdeuxcols]{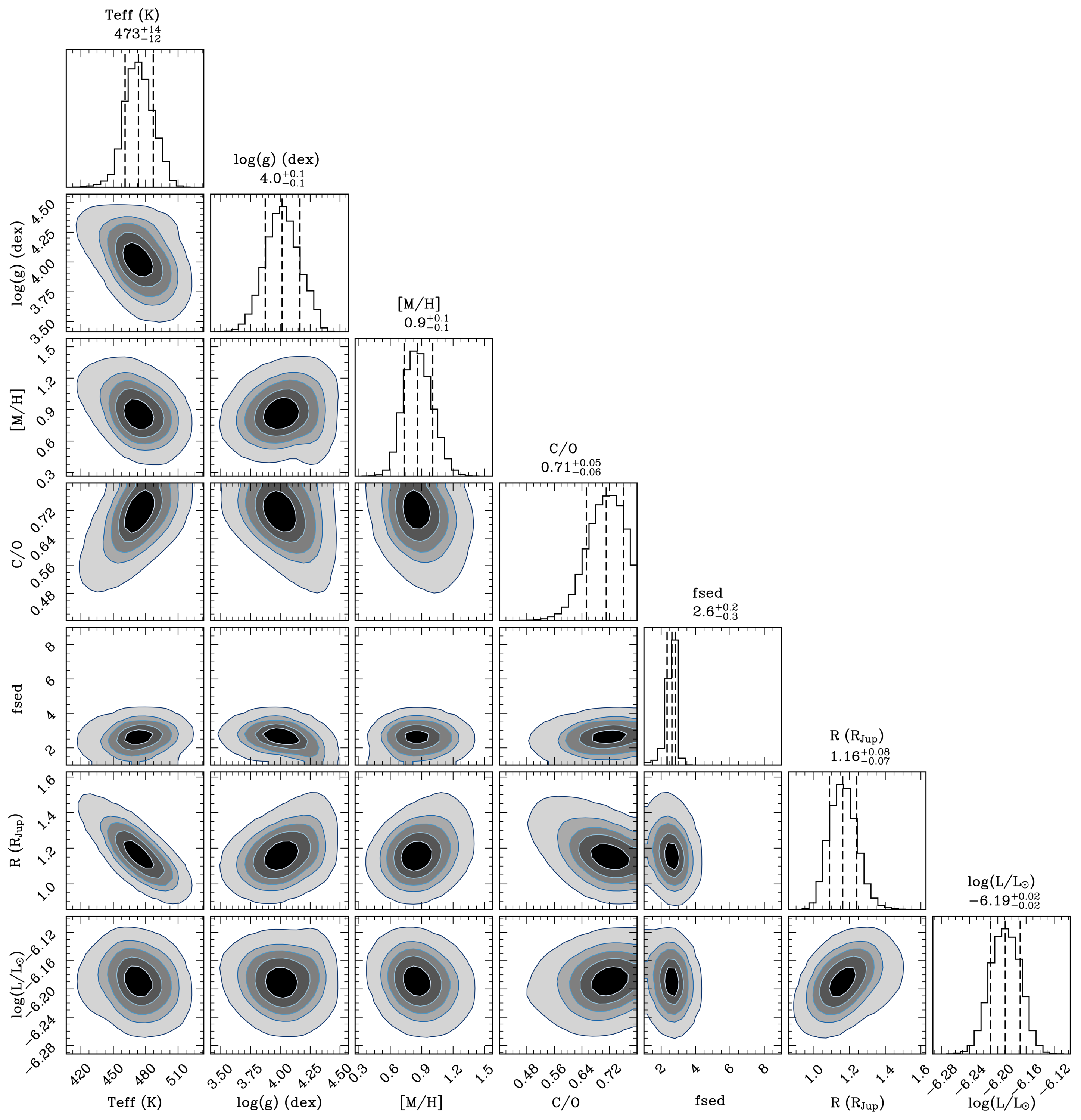}
    \caption{Posterior distribution from forward modelling with \texttt{ForMoSA} using the GJ~504\,b photometry with the low-resolution \exoII grid. The resulting best fit model is shown in Fig.~\ref{fig: bf_gj504b}.}
    \label{fig: gj504b corner}
\end{figure*}

In Figures~\ref{fig: vhs mrs corner}--\ref{fig: gj504b corner},
we provide the posterior distributions associated to each forward modelling analysis described in the main text.

\section{Longitudinal heterogeneity fraction calculation}
\label{subsection: longitudinal frac calc}
In Section~\ref{subsection: towards the mapping of clouds on vhs} we calculate the fraction of longitudinal heterogeneous cloud cover, using the fluxes from each of the \fsed solutions derived in the two-column approach. Here we detail the method. We suppose that, in the two-column framework, the maximum and minimum fluxes produced along a lightcurve are given by:
\begin{align}
F_{\mathrm{max}} &= \alpha_{\mathrm{max}} F_{\mathrm{thick}} + (1-\alpha_{\mathrm{max}}) F_{\mathrm{thin}} \\
F_{\mathrm{min}} &= \alpha_{\mathrm{min}} F_{\mathrm{thick}} + (1-\alpha_{\mathrm{min}}) F_{\mathrm{thin}} \notag\\
F_{\mathrm{max}} &= \langle F\rangle + \frac{\Delta F}{2} \notag\\
F_{\mathrm{min}} &= \langle F\rangle - \frac{\Delta F}{2}
\end{align}
with $\Delta F = F_{\mathrm{max}} - F_{\mathrm{min}}$.\\
\begin{align}
\Rightarrow \Delta \alpha = \frac{\Delta F}{F_{\mathrm{thick}}-F_{\mathrm{thin}}}
\end{align}
For the 2018 HST observed amplitude 1.2/4.71.of $\Aobs = 0.247 = \frac{\Delta F}{\langle F \rangle}$ at 1.27\,\mic, we have:
\begin{align}
F_{\mathrm{thick}} &= 3.1\times 10^{-16} \notag\\
F_{\mathrm{thin}} &= 7.8\times 10^{-16} \notag\\
F_{\mathrm{thin}} - F_{\mathrm{thick}} &= 4.7\times10^{-16}~\mathrm{W\,m^{-2}\,\mic^{-1}}
\end{align}
The average flux at 1.27\,\mic is taken to be the two-column best-fit solution that we find in Fig.~\ref{fig: vhs best fit} corresponding to $\alpha = 0.62$; $\langle F\rangle = 0.62F_{\mathrm{thick}} + 0.38F_{\mathrm{thin}} = 4.89\times10^{-16}~\mathrm{W\,m^{-2}\,\mic^{-1}}$.
The fractions of thick cloud coverage producing the minimum and maximum of the HST lightcurve are:
\begin{align}
\alpha_{\mathrm{max}} = \frac{F_{\mathrm{max}} - F_{\mathrm{thin}}}{F_{\mathrm{thick}} - F_{\mathrm{thin}}} = 0.50, \hspace{+0.3cm}\alpha_{\mathrm{min}} = \frac{F_{\mathrm{min}} - F_{\mathrm{thin}}}{F_{\mathrm{thick}} - F_{\mathrm{thin}}} = 0.75\notag
\end{align}
\begin{align}
\boxed{\therefore \Delta \alpha_{1.27} = 0.25}
\end{align}
Similarly, for the Spitzer observed amplitude of $\Aobs = 0.0576$ at 4.5\,\mic we have:
\begin{align}
F_{\mathrm{thick}} &= 2.7\times 10^{-16} \notag\\
F_{\mathrm{thin}} &= 8.0\times 10^{-17} \notag\\
F_{\mathrm{thin}}- F_{\mathrm{thick}} &= -1.9\times10^{-16}~\mathrm{W\,m^{-2}\,\mic^{-1}}
\end{align}
The average flux is $\langle F\rangle = 2.00\times10^{-16}~\mathrm{W\,m^{-2}\,\mic^{-1}}$.
Therefore, the fractions of thick cloud coverage producing the minimum and maximum of the Spitzer lightcurve are:
\begin{align}
\alpha_{\mathrm{max}} = \frac{F_{\mathrm{max}} - F_{\mathrm{thin}}}{F_{\mathrm{thick}} - F_{\mathrm{thin}}} = 0.65, \hspace{+0.3cm}\alpha_{\mathrm{min}} = \frac{F_{\mathrm{min}} - F_{\mathrm{thin}}}{F_{\mathrm{thick}} - F_{\mathrm{thin}}} = 0.59\notag
\end{align}
\begin{align}
\boxed{\therefore \Delta \alpha_{4.5} = 0.06}
\end{align}
\end{appendix}

\end{document}